\documentclass[fleqn,usenatbib,useAMS]{mnras}
\usepackage{graphicx,natbib,bm,url,color}
\graphicspath{{./fig/}{./png/}}
\newcommand{\EQ}{\begin{equation}}
\newcommand{\EN}{\end{equation}}
\newcommand{\EQA}{\begin{eqnarray}}
\newcommand{\ENA}{\end{eqnarray}}

\newcommand{\Eq}[1]{Eq.~(\ref{#1})}

\newcommand{\Eqss}[2]{Eqs.~(\ref{#1})--(\ref{#2})}

\newcommand{\App}[1]{Appendix~\ref{#1}}

\newcommand{\Sec}[1]{Sect.~\ref{#1}}

\newcommand{\Fig}[1]{Figure~\ref{#1}}

\newcommand{\Figp}[2]{Figure~\ref{#1}({#2})}
\newcommand{\Figsp}[3]{Figures~\ref{#1}({#2}) and ({#3})}

\newcommand{\Figs}[2]{Figures~\ref{#1} and \ref{#2}}

\newcommand{\Tab}[1]{Table~\ref{#1}}

\newcommand{\bra}[1]{\langle #1\rangle}
\newcommand{\bbra}[1]{\left\langle #1\right\rangle}

{}
{}
{}

{}
{}
{}
{}
{}
{}
{}
{}
{}
{}
{}
{}
{}
{}
{}
{}
{}

{}

{}

{}
{}
{}

\newcommand{\zzz}{\hat{\mbox{\boldmath $z$}} {}}

\newcommand{\kk}{\bm{k}}

\newcommand{\xx}{\bm{x}}

\newcommand{\BB}{\bm{B}}

\newcommand{\JJ}{\bm{J}}
\newcommand{\oo}{\bm{\omega}}

\newcommand{\AAA}{\bm{A}}

\newcommand{\uu}{\bm{u}}

\newcommand{\nab}{{\bm{\nabla}}}

\newcommand{\SSSS}{\mbox{\boldmath ${\sf S}$} {}}

\newcommand{\ii}{{\rm i}}

\newcommand{\dive}{{\rm div}  \, {}}

\newcommand{\DD}{{\rm D} {}}

\newcommand{\dd}{{\rm d} {}}

\def\Ma{\mbox{\rm Ma}}

\def\Pm{\mbox{\rm Pr}_{\rm M}}
\def\Rm{\mbox{\rm Re}_{\rm M}}

\def\Rmc{\mbox{\rm Re}_{\rm M}^{\rm crit}}

\def\Rey{\mbox{\rm Re}}

\def\Lu{\mbox{\rm Lu}}

\def\EEP{{\cal E}_{\rm P}}
\def\EEK{{\cal E}_{\rm K}}
\def\EEM{{\cal E}_{\rm M}}

\def\EK{E_{\rm K}}
\def\EM{E_{\rm M}}

\def\cs{c_{\rm s}}

\def\vA{v_{\rm A}}

\def\kf{k_{\rm f}}

\def\HM{H_{\rm M}}
\def\HK{H_{\rm K}}
\def\EM{E_{\rm M}}

\def\Brms{B_{\rm rms}}

\def\urms{u_{\rm rms}}

\def\orms{\omega_{\rm rms}}

\def\half{{\textstyle{1\over2}}}

\def\onethird{{\textstyle{1\over3}}}

\newcommand{\g}{\,{\rm g}}
\newcommand{\s}{\,{\rm s}}

\newcommand{\cm}{\,{\rm cm}}

\newcommand{\km}{\,{\rm km}}

\newcommand{\kms}{\,{\rm km\,s}^{-1}}

\newcommand{\pc}{\,{\rm pc}}

\newcommand{\Myr}{\,{\rm Myr}}

\newcommand{\erg}{\,{\rm erg}}

\def\apj{ApJ}
\def\apjl{ApJL}
\def\apjs{ApJS}
\def\prd{PhRvD}
\def\pre{PhRvE}
\def\prf{PhRvF}
\def\prl{PhRvL}

\def\araa{ARA\&A}

\def\aap{A\&A}
\def\mnras{MNRAS}
\def\aapr{A\&AR}

\def\nat{Natur}

\definecolor{battleshipgrey}{rgb}{0.52, 0.52, 0.51}
\definecolor{upforestgreen}{rgb}{0.0, 0.27, 0.13}
\definecolor{ao(english)}{rgb}{0.0, 0.5, 0.0}
\definecolor{airforceblue}{rgb}{0.36, 0.54, 0.66}
\definecolor{aurometalsaurus}{rgb}{0.43, 0.5, 0.5}
\definecolor{amber(sae/ece)}{rgb}{1.0, 0.49, 0.0}
\definecolor{azure(colorwheel)}{rgb}{0.0, 0.5, 1.0}
\definecolor{britishracinggreen}{rgb}{0.0, 0.26, 0.15}

\title{Dynamo effect in unstirred self-gravitating turbulence}
\author[A. Brandenburg \& E. Ntormousi]{Axel Brandenburg$^{1,2,3,4}$\thanks{E-mail:brandenb@nordita.org}
 and Evangelia Ntormousi$^{5,6}$
\\
$^1$Nordita, KTH Royal Institute of Technology and Stockholm University, Hannes Alfv\'ens v\"ag 12, SE-10691 Stockholm, Sweden \\
$^2$The Oskar Klein Centre, Department of Astronomy, Stockholm University, AlbaNova, SE-10691 Stockholm, Sweden\\
$^3$McWilliams Center for Cosmology and Department of Physics, Carnegie Mellon University, 5000 Forbes Ave, Pittsburgh, PA 15213, USA\\
$^4$School of Natural Sciences and Medicine, Ilia State University, 3-5 Cholokashvili Avenue, 0194 Tbilisi, Georgia\\
$^5$Scuola Normale Superiore, Piazza dei Cavalieri 7, I-56126 Pisa, Italy \\
$^6$Institute of Astrophysics, Foundation for Research and Technology-Hellas, Vasilika Vouton, GR-70013 Heraklion, Greece
}

\begin{document}
\maketitle

\begin{abstract}
In many astrophysical environments, self-gravity can generate kinetic
energy, which, in principle, is available for driving dynamo action.
Using direct numerical simulations, we show that in unstirred
self-gravitating subsonic turbulence with helicity and a magnetic
Prandtl number of unity, there is a critical magnetic Reynolds number
of about 25 above which the work done against the Lorentz force exceeds
the Ohmic dissipation.
The collapse itself drives predominantly irrotational motions that
cannot be responsible for dynamo action.
We find that, with a weak magnetic field, one-third of the work done
by the gravitational force goes into compressional heating and the
remaining two-thirds go first into kinetic energy of the turbulence
before a fraction of it is converted further into magnetic and finally
thermal energies.
Close to the collapse, however, these fractions change toward 1/4 and
3/4 for compressional heating and kinetic energy, respectively.
When the magnetic field is strong, the compressional
heating fraction is unchanged.
Out of the remaining kinetic energy, one quarter goes directly into
magnetic energy via work against the Lorentz force.
The fraction of vortical motions diminishes in favor of compressive
motions that are almost exclusively driven by the Jeans instability.
For an initially uniform magnetic field, field amplification at scales
larger than those of the initial turbulence are driven by tangling.
\end{abstract}


\section{Introduction}

Dynamo action describes the conversion of kinetic energy into magnetic
\citep{Mof78}.
This can also happen under non-stationary conditions, for example
in decaying turbulence, where the kinetic energy tends to decay
in power-law fashion, so the growth of the magnetic field is no
longer exponential in time, as it would be in stationary turbulence
\citep{Bran+19,Sur2019}.
This type of unsteady energy conversion is expected to play a role in many
astrophysical settings where the magnetic Reynolds number is high enough.

In the interstellar medium (ISM), as well as on cosmological scales, self-gravity
can be the dominant driver of turbulence \citep{field_2008,klessen_hennebelle_2010}.
In the context of galaxy growth, the gas accretion flows typically
predicted by numerical simulations around galactic disks \citep[for
some examples]{dekel2009,nelson2015} convert gravitational
potential energy into kinetic energy.
In the ISM of galaxies, the gravitational instability of the
disk itself has been proposed as the main source of turbulence
\citep[e.g.,][]{Bournaud2010,Krumholz_Burkhart_2016}, with
a contribution that can be as strong as that of supernova (SN)
feedback \citep{krumholtz+18}. Finally, gravitational accretion is
believed to be the main driver of turbulence within molecular clouds
\citep{Ibanez-Mejia2017}.

While the exact fraction of potential energy that goes into turbulence
is unknown -- and probably depends strongly on the environment -- it
is clear that, in general, self-gravity can be an important source of
kinetic energy.
This kinetic energy can, in turn, be converted into magnetic energy
through dynamo action.
Earlier work has shown that in gravitationally unstable flows,
the magnetic energy increases during the linear phase of
the collapse, and that the magnetic energy declines during the
nonlinear phase of the dynamo \citep{Sur+10,Sur+12,XL20}.

The kinematic phase of a turbulent dynamo within a collapsing cloud was
considered by \cite{Federrath+11b}, who reported exponential growth of
the magnetic field with a Kazantsev spectrum, as was previously found
for forced turbulence.
Their dynamo growth rate exceeded the collapse rate, provided their
resolution criterion of 30 grid cells per Jeans length is obeyed at any
position in space and any point in time.
As they demonstrated, at lower resolution, the magnetic field is only amplified 
because of flux freezing.

Magnetic fields can also be produced by tangling of a large-scale seed.
This type of growth can still occur in two dimensions, where true
dynamo action is impossible according to the Cowling antidynamo theorem;
see \cite{Hide+Palmer82} for a generalized antidynamo theorem relevant
to compressible flows.

An important goal of this work is to characterize dynamo action in
unstirred decaying turbulence, where gravity provides an energy source
that can eventually revert the decay of the turbulence. 
We approach the question with direct numerical simulations of systems
with different turbulent initial conditions, and in various degrees of
gravitational instability.
Our turbulent initial conditions are almost exclusively subsonic with
a Mach number of 0.2.
One motivation behind this choice is that the critical magnetic Reynolds
number for dynamo action increases by about a factor of two when the
flow is supersonic \citep{Hau+04b}.
The low initial Mach number also allows us to focus on any dynamo action
triggered by the collapse-produced turbulence, rather than by the decaying
turbulence from the initial conditions.
In fact, we will show that, as the models evolve, the collapse itself
produces turbulence that eventually dominates the initial flow.
However, since turbulence in many astrophysical environments, such as
molecular clouds, is supersonic \citep{Schneider+13}, we will also
present one run with an initial Mach number of two.
Furthermore, given that the sonic Mach number is scale-dependent 
\citep{Federrath+21}, our early subsonic phase might still
be applicable to correspondingly small scales.

In driven turbulence, dynamo action can be adequately characterized by
the growth rate, evaluated as the time derivative of the root-mean-square
(rms) magnetic field.
In stationary conditions, this quantity stays reasonably constant with time.
However, in the non-stationary conditions that we study here, namely
decaying turbulence and turbulence generated by gravitational collapse,
the magnetic field no longer grows exponentially, and
the growth rate cannot be used as a dynamo criterion.
Therefore, in this study we explore new, more general dynamo criteria
that allow for non-stationary turbulence conditions.
We decided to base our dynamo criterion on the work against the Lorentz
force, where the magnetic curvature force plays the dominant role.
When this work exceeds the Joule dissipation, it might be a dynamo.
This definition of a possible dynamo agrees with the standard definition
of a positive growth rate when the flow is steady, but, unlike any dynamo
criterion proposed so far, it can easily be applied to unsteady flows.
It does not, however, distinguish dynamos in three dimensions from just
temporary amplification through tangling and compression, as can be seen,
for example, in two dimensions.
To exclude the effects of two-dimensional (2-D) compression or tangling,
we propose splitting the Lorentz work term into two contributions,
of which one is absent in 2-D.
This leads to an additional criterion that must be satisfied for dynamo
action.

We use high-resolution numerical simulations with fixed kinematic
viscosity and magnetic diffusivity to be able to define the threshold
for dynamo action.
Note also that, unlike codes with adaptive mesh refinement, where
the accuracy of the solution varies in space \citep[see, for
example,][]{Federrath+10}, we resolve all regions in space equally well.
There should therefore be no doubt that our velocity spectra and other
diagnostics are representative of the domain as a whole.

In this paper, we first define our model (\Sec{TheModel}).
We then present the results for the energy spectra and the energy
conversion rates, as well as characteristic wavenumbers and dynamo
excitation conditions for weak initial magnetic fields (\Sec{Results}).
We then compare our results with those for strong initial magnetic fields
(\Sec{ResultsStrong}), and conclude in \Sec{Conclusions}.

\section{The model}
\label{TheModel}

\subsection{Governing equations}
\label{GoverningEquations}

We consider an isothermal gas with sound speed $\cs$ in a
cubic periodic domain of size $L^3$, so the smallest wavenumber
is $k_1=2\pi/L$.
The pressure is given by $p=\rho\cs^2$, where $\rho$ is the density.
The governing equations are \citep{Passot+95}
\begin{equation}
\nabla^2\Phi=4\pi G\left(\rho-\rho_0\right),
\label{del2Phi}
\end{equation}
\begin{equation}
{\DD\uu\over\DD t}=-\nab\left(\cs^2\ln\rho+\Phi\right)
+\frac{1}{\rho}\left(\JJ\times\BB+\nab\cdot2\rho\nu\SSSS\right),
\end{equation}
\begin{equation}
{\DD\ln\rho\over\DD t}=-\nab\cdot\uu,
\label{Dlnrho}
\end{equation}
\begin{equation}
{\partial\AAA\over\partial t}=\uu\times\BB-\eta\mu_0\JJ,
\label{dAdt}
\end{equation}
where $\Phi$ is the gravitational potential,
$G$ is Newton's constant, $\rho_0$ is the spatially averaged density,
which is constant in time because of mass conservation,
$\uu$ is the velocity, $\JJ=\nab\times\BB/\mu_0$ is the current density,
$\mu_0$ is the vacuum permeability, $\BB=\nab\times\AAA$ is
the magnetic field in terms of the magnetic vector potential,
${\sf S}_{ij}=\half(u_{i,j}+u_{j,i})-\onethird\delta_{ij}\nab\cdot\uu$
are the components of the rate-of-strain tensor with commas denoting
partial derivatives, and $\nu$ is the kinematic viscosity.

The occurrence of the constant $\rho_0$ in \Eq{del2Phi} comes from a
change of coordinates to a comoving reference frame that is following the global
expansion of the background medium \citep[see][]{AL88}.
This is a consequence of working with an infinite (unbounded) medium.
Such a medium can be stationary, but not static.
This was also explained by \cite{Falco+13}, who clarified why the
famous Jeans swindle \citep{BT08} actually works.
In this accelerated frame, \Eqss{del2Phi}{dAdt} describe the departure
of collapsing structures from the background flow.

Linearizing \Eqss{del2Phi}{Dlnrho} around $\rho=\rho_0$ and $\uu=0$, and
assuming the perturbations to be proportional to $e^{\ii\kk\cdot\xx+\sigma t}$
yields the dispersion relation $\sigma^2=\sigma_{\rm J}^2-\cs^2 k^2$,
where $\sigma_{\rm J}^2=4\pi G\rho_0$ with $\sigma_{\rm J}$ being the
gravitational or Jeans growth rate \citep{Jeans02}.
The pressureless free-fall time is $t_{\rm ff}=\sqrt{3/8}\,\pi/\sigma_{\rm J}
\approx1.92/\sigma_{\rm J}$ \citep{Shu92}.
The Jeans wavenumber is $k_{\rm J}=\sigma_{\rm J}/\cs$, and the
Jeans length is then $\lambda_{\rm J}=2\pi/k_{\rm J}$
\citep[see, e.g.,][]{Bonazzola+87,Tru+97}.
According to the classical Jeans criterion for gravitational instability,
an interstellar gas cloud will collapse if its free-fall time is shorter
than the sound crossing time in its interior, or, more specifically,
$t_{\rm ff}\cs k_1<1.92$.

\subsection{Diagnostic quantities}

\subsubsection{Energetics}
\label{Energetics}

Throughout this paper, we use periodic boundary conditions, so all
surface integrals vanish and no mass is lost.
In the following, volume averages are denoted by angle brackets.
It is instructive to inspect the evolution equations of mean potential,
kinetic, and magnetic energy densities, $\EEP=-\bra{(\nab\Phi)^2}/8\pi G$,
$\EEK=\bra{\rho_0\uu^2}/2$, and $\EEM=\bra{\BB^2}/2\mu_0$, respectively
\citep{Banerjee+Kritsuk18}.
They are given by
\begin{equation}
\frac{\dd\EEP}{\dd t}=-W_{\rm J},
\label{dEP}
\end{equation}
\begin{equation}
\frac{\dd\EEK}{\dd t}=W_{\rm P}+W_{\rm J}+W_{\rm L}-Q_{\rm K},
\label{dEK}
\end{equation}
\begin{equation}
\frac{\dd\EEM}{\dd t}=-W_{\rm L}-Q_{\rm M},
\end{equation}
where $W_{\rm P}=-\bra{\uu\cdot\nab p}=\bra{p\nab\cdot\uu}$ is the work
done by the pressure force, $W_{\rm J}=-\bra{\rho\uu\cdot\nab\Phi}$ is the
work done by the gravity term, $W_{\rm L}=\bra{\uu\cdot(\JJ\times\BB)}$ is
the work done by the Lorentz force, and $Q_{\rm K}=\bra{2\rho\nu\SSSS^2}$
and $Q_{\rm M}=\bra{\mu_0\eta\JJ^2}$ are the viscous and Joule dissipation
terms.
The thermal energy density is sourced by the terms
$-W_{\rm P}+Q_{\rm K}+Q_{\rm M}$, but with the employed isothermal
equation of state, the thermal energy density is not evolved.

The work done by the gravity term ($W_{\rm J}>0$) leads to a decrease of the
potential energy density and to an increase in the kinetic energy density.
During the collapse, the virial parameter
$\alpha_{\rm vir}=2\EEK/|{\cal E}_{\rm P}|$ is expected to be around
unity, but this expectation can be different at large Mach numbers
\citep{NH2019} and for strong magnetic fields; see \cite{Federrath+12},
who also emphasize the difference between periodic setups, such as ours,
and isolated spheres.

We have defined all work terms such that they enter with a plus sign in
\Eq{dEK}, i.e., they lead to an increase in the kinetic energy density
if they are positive, and thus to a loss in some other energy reservoir.
As noted above, a positive $W_{\rm J}$ term leads to a loss of potential
energy.
Likewise, a positive $W_{\rm P}$ term leads to a loss in thermal energy.
Gravitational collapse however, leads to compressional heating and
$W_{\rm P}$ is therefore negative.
Furthermore, dynamo action leads to a growth in magnetic energy if 
$W_{\rm L}$ is negative.

The $W_{\rm L}$ term can be split into three constituents:
$W_{\rm L}^{\rm c}=-\bra{\uu\cdot\nab\BB^2/2\mu_0}$,
$W_{\rm L}^\|=\bra{\uu\cdot(\BB\cdot\nab\BB/\mu_0)_\|}$, and
$W_{\rm L}^\perp=\bra{\uu\cdot(\BB\cdot\nab\BB/\mu_0)_\perp}$.
Here, $-\nab\BB^2/2\mu_0$ is the magnetic pressure contribution
of $\JJ\times\BB$, and $(\BB\cdot\nab\BB/\mu_0)_\|$ and
$(\BB\cdot\nab\BB/\mu_0)_\perp$ are the stretching terms along
and perpendicular to the magnetic field.
The last two forces are also referred to as tension and curvature forces;
see \cite{Nor+92} for their contributions to a convective dynamo.

In the following, we also decompose $W_{\rm L}$ by writing it as
$W_{\rm L}=-\bra{\JJ\cdot(\uu\times\BB)}$ and expanding the curl to get
\EQ
-\bra{\JJ\cdot(\uu\times\BB)}=\bra{J_i u_j (A_{i,j}-A_{j,i}}
\equiv W_{\rm L}^{\rm2D}+W_{\rm L}^{\rm3D}.
\EN
Here, we make use of the fact that the Weyl gauge has been used in
\Eq{dAdt}.
In two dimensions, the magnetic field can be represented as
$\BB=\nab\times A_z\zzz$, with its $x$ and $y$ components lying in the
$xy$ plane.
Then the term $W_{\rm L}^{\rm 3D}=-\bra{J_i u_j A_{j,i}}$ vanishes in 2-D.
Thus, we can identify $W_{\rm L}^{\rm3D}$ with a contribution that
characterizes the 3-D nature of the system and can therefore be a proxy
for dynamo action, provided $W_{\rm L}^{\rm3D}$ is large enough.

To characterize the flow of energy, it is convenient to define the
fractions $\epsilon_{\rm J}^{\rm P}\equiv -W_{\rm P}/W_{\rm J}$,
$\epsilon_{\rm J}^{\rm L}\equiv -W_{\rm L}/W_{\rm J}$,
$\epsilon_{\rm J}^{\rm+K}\equiv\dot{\cal E}_{\rm K}/W_{\rm J}$,
and $\epsilon_{\rm J}^{\rm-K}=Q_{\rm K}/W_{\rm J}$.
Likewise, we define the fractions
$\epsilon_{\rm L}^{\rm+M}\equiv\dot{\cal E}_{\rm M}/(-W_{\rm L})$,
and $\epsilon_{\rm J}^{\rm-M}=Q_{\rm M}/(-W_{\rm L})$.
To characterize the growth or decay of the magnetic field, we define
the nondimensional ratio $\epsilon_{\rm M}^\Delta=(-W_{\rm L}-Q_{\rm M})/Q_{\rm M}$.
A related quantity is the pseudo (or instantaneous) growth rate of magnetic energy,
$\gamma=(-W_{\rm L}-Q_{\rm M})/\EEM$, which can be divided into the
contributions $\gamma_{\rm c}+\gamma_\|+\gamma_\perp=\gamma$ from compression
and stretching parallel and perpendicular to the magnetic field, where
$\gamma_\perp=(-W_{\rm L}^\perp-Q_{\rm M})/E_{\rm M}$ will play the
most important role, and $\gamma_{\rm c}=-W_{\rm L}^\|/E_{\rm M}$ and
$\gamma_\|=-W_{\rm L}^{\rm c}/E_{\rm M}$ contribute either later or in
the presence of strong initial magnetic fields.
Likewise, we define $\gamma_{\rm2D}=-(W_{\rm L}^{\rm2D}+Q_{\rm M})/E_{\rm M}$
and $\gamma_{\rm3D}=-W_{\rm L}^{\rm3D}/E_{\rm M}$, so that
$\gamma_{\rm2D}+\gamma_{\rm3D}=\gamma$.

\subsubsection{Characteristic wavenumbers}

To characterize the compressive and solenoidal flow components,
it is convenient to compute the rms velocity divergence,
$(\nab\cdot\uu)_{\rm rms}=\bbra{(\nab\cdot\uu)^2}^{1/2}$,
and the rms vorticity, $\orms=\bra{\oo^2}^{1/2}$, where
$\oo=\nab\times\uu$, and to define
\EQ
k_{\,\nab\cdot\uu}=(\nab\cdot\uu)_{\rm rms}/\urms,
\EN
\EQ
k_{\oo}=\orms/\urms,
\EN
which have the dimension of a wavenumber.
Since the flow is helical, we can also define the wavenumber
\EQ
k_{\oo\cdot\uu}=|\bra{\oo\cdot\uu}|/\urms^2,
\EN
which characterizes the typical wavenumber where helicity plays a role.
Large values of $k_{\,\nab\cdot\uu}$, $k_{\oo}$, and $k_{\oo\cdot\uu}$
imply strong flow divergences or compressions, strong vortices, and
strong swirls, respectively.
To characterize the flow compression from the gravitational collapse,
we also define
\EQ
k_{p\nab\cdot\uu}=-\bra{p\nab\cdot\uu}/p_0\urms\quad
\mbox{(when $k_{p\nab\cdot\uu}>0$)},
\EN
where $p_0=\rho_0\cs^2$ has been introduced for brevity.
The relevance of $k_{p\nab\cdot\uu}$ to the collapse phenomenon is
motivated by the fact that a strong flow compression or flow convergence
($\nab\cdot\uu<0$) correlates with pressure (defined in the
beginning of \Sec{GoverningEquations}).
Indeed, it turns out that $k_{p\nab\cdot\uu}$ is very small prior to
collapse, but it approaches $k_{\nab\cdot\uu}$ close to the collapse.

\subsubsection{Spectra}

We define the kinetic and magnetic energy spectra, $\EK(k,t)$ and
$\EM(k,t)$, respectively.
They are normalized such that $\int\EK(k,t)\,\dd k=\EEK$ and
$\int\EM(k,t)\,\dd k=\EEM$.
It can be advantageous to express them as wavenumber-dependent Reynolds
and Lundquist numbers by defining a velocity and a magnetic field,
\EQ
u_k(t)=\sqrt{2k\EK(k,t)/\rho_0},\quad
B_k(t)=\sqrt{2\mu_0 k\EM(k,t)},
\EN
respectively.
We then define
\EQ
\Rey_k(t)=u_k(t)/\nu k\quad\mbox{and}\quad
\Lu_k(t)=B_k(t)/(\sqrt{\mu_0\rho_0}\,\eta k).
\EN
A Kolmogorov-type spectrum with $\EK(k)\propto k^{-5/3}$ corresponds
then to $u_k\propto k^{-1/3}$ and $\Rey_k\propto k^{-4/3}$.
In the following, we also quote these values at $k=\kf$, where
the initial kinetic energy spectrum peaks.
We also define the Reynolds number $\Rey_t$ based on the actual rms
velocity, with $\Rey_{t_*}$ denoting the value at the time $t_*$, when
the exponentially growing gas motions from the Jeans instability begin
to dominate over the initial turbulence.
In the present normalization, the values of $\Rey_{\kf}$ and $\Rey_t$
are close to the Taylor microscale Reynolds number \citep{TL72}, which
is universally defined as $\Rey_\lambda=v'\lambda_{\rm Tay}/\nu$.
Here, $v'=\urms/\sqrt{3}$ is the one-dimensional rms velocity and
$\lambda_{\rm Tay}=\sqrt{15\nu\rho_0/Q_{\rm K}}\,v'$ is the Taylor
microscale.

Kinetic and magnetic helicity spectra, $\HK(k,t)$ and $\HM(k,t)$, are
normalized such that $\int \HK(k,t)\,\dd k=\bra{\oo\cdot\uu}$ and 
$\int \HM(k,t)\,\dd k=\bra{\AAA\cdot\BB}$ are the mean kinetic and
magnetic helicities.
They obey the realizability conditions $|\HK(k)|\leq2k\EK(k)$ and
$|\HM(k)|\leq(2/k)\EM(k)$ \citep{Mof78}.
It is then convenient to plot the relative helicities given by
the corresponding ratios $\HK(k)/2k\EK(k)$ and $k\HM(k)/2\EM(k)$,
respectively.

We also plot the enstrophy and logarithmic density spectra, $E_\omega(k)$
and $E_{\ln\rho}(k)$, respectively.
They are normalized such that $\int E_\omega(k)\,\dd k=\bra{\oo^2}/2$
and $\int E_{\ln\rho}(k)\,\dd k=\bra{(\ln\rho)^2}$.
Here, $E_\omega(k)/k^2$ corresponds to the kinetic energy spectrum
of the vortical part of the velocity, and $E_\rho(k)$ reflects its
irrotational part.
Finally, the potential energy spectrum is normalized such that
$\int E_{\rm P}(k)\,\dd k=\EEP$.

\subsection{Units and parameters}
\label{Units}

In the plots shown in this paper, we express velocities in units of $\cs$,
lengths in units of $k_1^{-1}$ (defined in the beginning of
\Sec{GoverningEquations}), density in units of $\rho_0$,
and the magnetic field in units of $\sqrt{\mu_0\rho_0}\cs$.
In practice, we do this by choosing in our simulations $\cs=k_1=\rho_0=\mu_0=1$.
In the following, to remind the readers, we often include relevant
combinations of $\cs$ and $k_1$ when specifying the numerical values,
but in other cases, especially in the table entries, we simply omit
them to avoid lengthy notation.

Since we do not invoke cooling or any other processes that depend on
dimensions, our simulations can be scaled to any arbitrary system
by choosing physical values for $\cs$, $k_1$, and $\rho_0$.
To illustrate the normalization used in the simulations, let us
consider, as an example, the case $\cs=1\kms$ and $k_1=1\pc^{-1}$.
Then, our time unit is $(\cs k_1)^{-1}=0.98\Myr$, so we can think
of our normalized time as $1\Myr$.
Considering a typical density of $\rho_0=10^{-21}\g\cm^{-3}$ for the
dense regions of the ISM, we have $\sigma_{\rm J}^{-1}=1.1\Myr$, or
$t_{\rm ff}=2.1\Myr$.
Then, the corresponding Jeans wavenumber is $k_{\rm J}=0.9\pc^{-1}$,
so the Jeans length is $\lambda_{\rm J}=7\pc$.
For $k_1=1\pc^{-1}$, the side length of the computational domain is
$6.28\pc\approx0.9\lambda_{\rm J}$.
The corresponding normalized (nondimensional) quantities are then
$\sigma_{\rm J}^{-1}\cs k_1=1.1$ and
$\sigma_{\rm J}/\cs k_1=k_{\rm J}/k_1=0.9$.
All work terms are given in units of $\rho_0\cs^3 k_1$, which corresponds
to $1\g\cm^{-3}(\km\s^{-1})^3/\pc=10^{-11}\erg\cm^{-3}\Myr^{-1}$ or
$0.0024\,L_\odot\pc^{-3}$.
In this work, we choose two values for $\sigma_{\rm J}/\cs k_1$, 2 and 5,
which means that our computational domain is two or five Jeans lengths
long, and our mesh of $2048^3$ cells resolves the Jeans length initially
with 1024 or 410 points, respectively.
As the collapse proceeds, the maximum density increases and the
nominal Jeans length decreases.
To stay within the resolution criterion of \cite{Federrath+11b} of
30 mesh points per Jeans length, we can only trust the time before
the maximum density has exceeded the initial value by a factor of
$(1024/30)^2\approx1200$ and $(410/30)^2\approx200$ for the cases
with $\sigma_{\rm J}/\cs k_1=2$ and 5, respectively.

As mentioned earlier, we focus on subsonic turbulence, where dynamo
action is most easily obtained \citep{Hau+04b,Federrath+11}.
In the context of molecular cloud contraction, this choice puts
them in the regime of low-mass pre-stellar cores.
While molecular clouds are supersonically turbulent on
large scales, low-mass pre-stellar cores are subsonic
\citep{larson_1981,Myers1991,Andre2007}.
Such weak motions could originate from the decay of larger-scale
turbulence \citep[e.g.,][]{Henn_falg_rev2012}.

We choose the amplitude of the initial velocity field such that the
initial Mach number $\Ma=\urms/\cs$ is around 0.1.
The turnover time is given by $\tau=(\urms\kf)^{-1}$.
We are interested in the cases where the turnover time is comparable
to or less than the free-fall time scale $\sigma_{\rm J}^{-1}$, where
$\sigma_{\rm J}$ must be larger than unity (in units of $\cs k_1$)
for the Jeans instability to be excited.
Given that $\Ma\approx0.1$, this automatically implies that
$\kf/k_1\geq10$, provided that $\sigma_{\rm J}$ is not much larger
than unity.
We focus on the case with $\sigma_{\rm J}=5\cs k_1$, but we have
also experimented with smaller values of two and even 1.1.
However, when $\sigma_{\rm J}$ is that small, it takes a long time
for the instability to develop and by that time the initial
turbulence would have decayed too much.

The magnetic field strength can also be specified in terms of
$\vA^{\rm rms}=\Brms/\sqrt{\mu_0\rho_0}$.
In the second part of the paper, we consider values of
$\vA^{\rm rms}/\cs$ in the range 0.04--0.4.
In the first part of the paper, however, we are interested in the
kinematic regime and therefore consider values of around $10^{-18}$.
In all cases, we adopt a magnetic Prandtl number of unity,
i.e., $\nu/\eta=1$, so the Reynolds number is always equal to
the magnetic Reynolds number.

\subsection{Initial conditions}

As initial conditions, we assume $\rho=\rho_0$, so there is no density
perturbation.
However, we assume that the velocity and magnetic fields have a random
distribution with a $k^4$ spectrum below a given wavenumber $\kf$ and
a $k^{-5/3}$ spectrum above $\kf$.
We assume the initial velocity to be maximally helical at all
wavenumbers, but take the magnetic field to be nonhelical.
This then leads to perturbations in the system that trigger the Jeans
instability.
Again, with a few exceptions, we deliberately choose a very weak initial
magnetic field so as to see the possibility of a kinematic dynamo at
early times.
A dynamo effect in decaying turbulence has been found in an earlier
study \citep{Bran+19}, where one saw a significant temporal increase of the
magnetic field over several orders of magnitude when the initial field
is sufficiently weak, 
but no significant increase was found for fields that start off
in near-equipartition with the turbulence.

\begin{figure}\begin{center}
\includegraphics[width=\columnwidth]{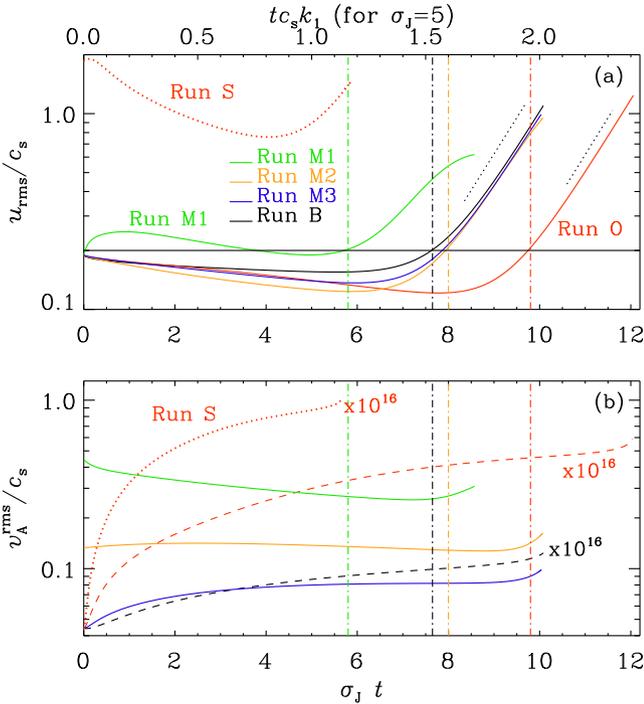}
\end{center}\caption[]{
Mach numbers for (a) the rms velocity and (b) the rms Alfv\'en speed.
The red and black dashed lines in (b) have been scaled up by
$10^{16}$ to make them visible.
The vertical dashed-dotted lines denote the times $t_*$ when the
Mach number has recovered to the original value of about 0.2.
(For Run~S, the initial value of 2 was not recovered before the collapse.)
The dotted lines correspond to $\exp(\sigma t)$ with
$\sigma=0.9\sqrt{24}$ and $\sqrt{3}$, respectively.
The upper abscissa gives time as $t\cs k_1$ for the cases with
$\sigma_{\rm J}/\cs k_1=5$.
}\label{pcomp_urms2}\end{figure}

\begin{figure*}\begin{center}
\includegraphics[width=.49\textwidth]{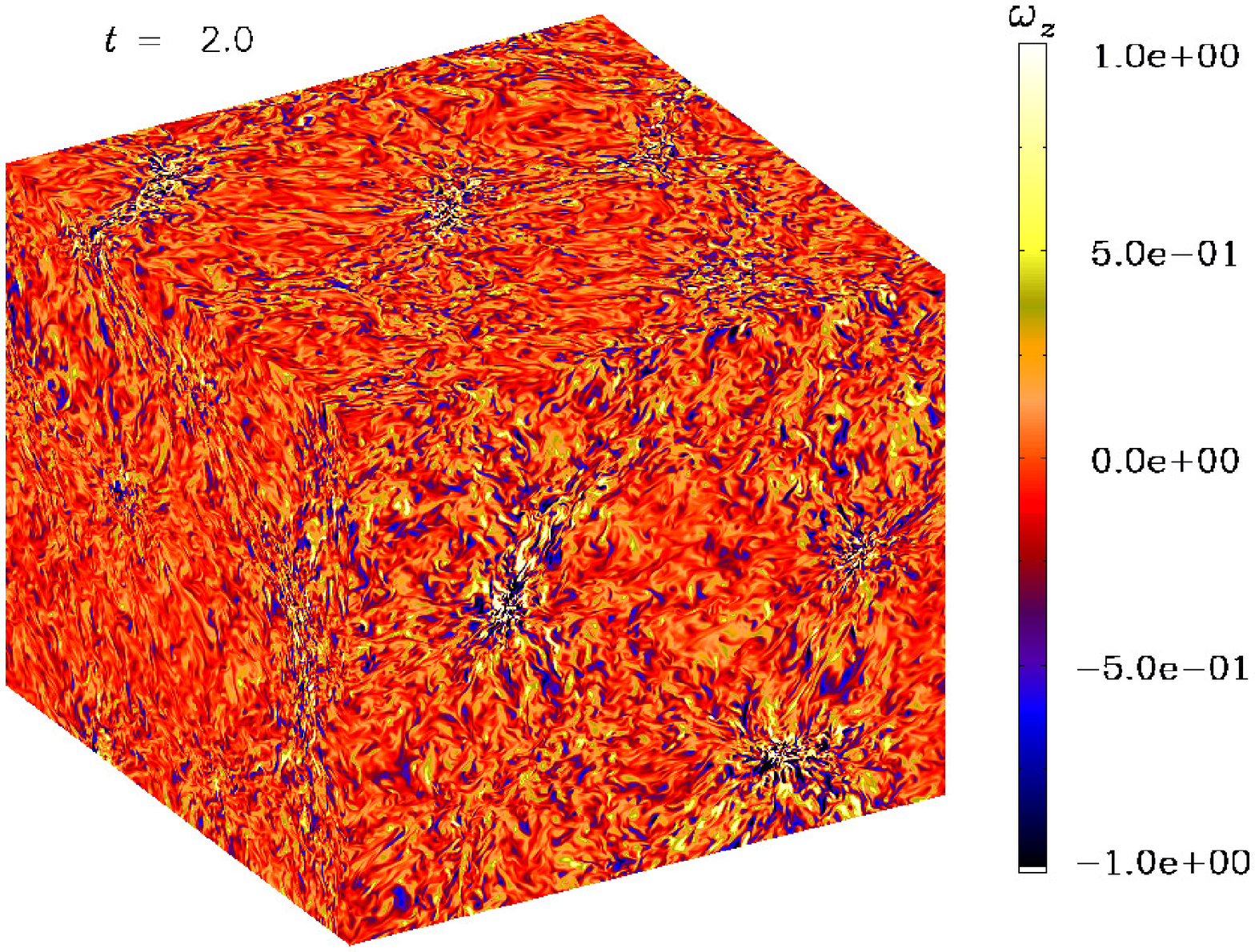}
\includegraphics[width=.49\textwidth]{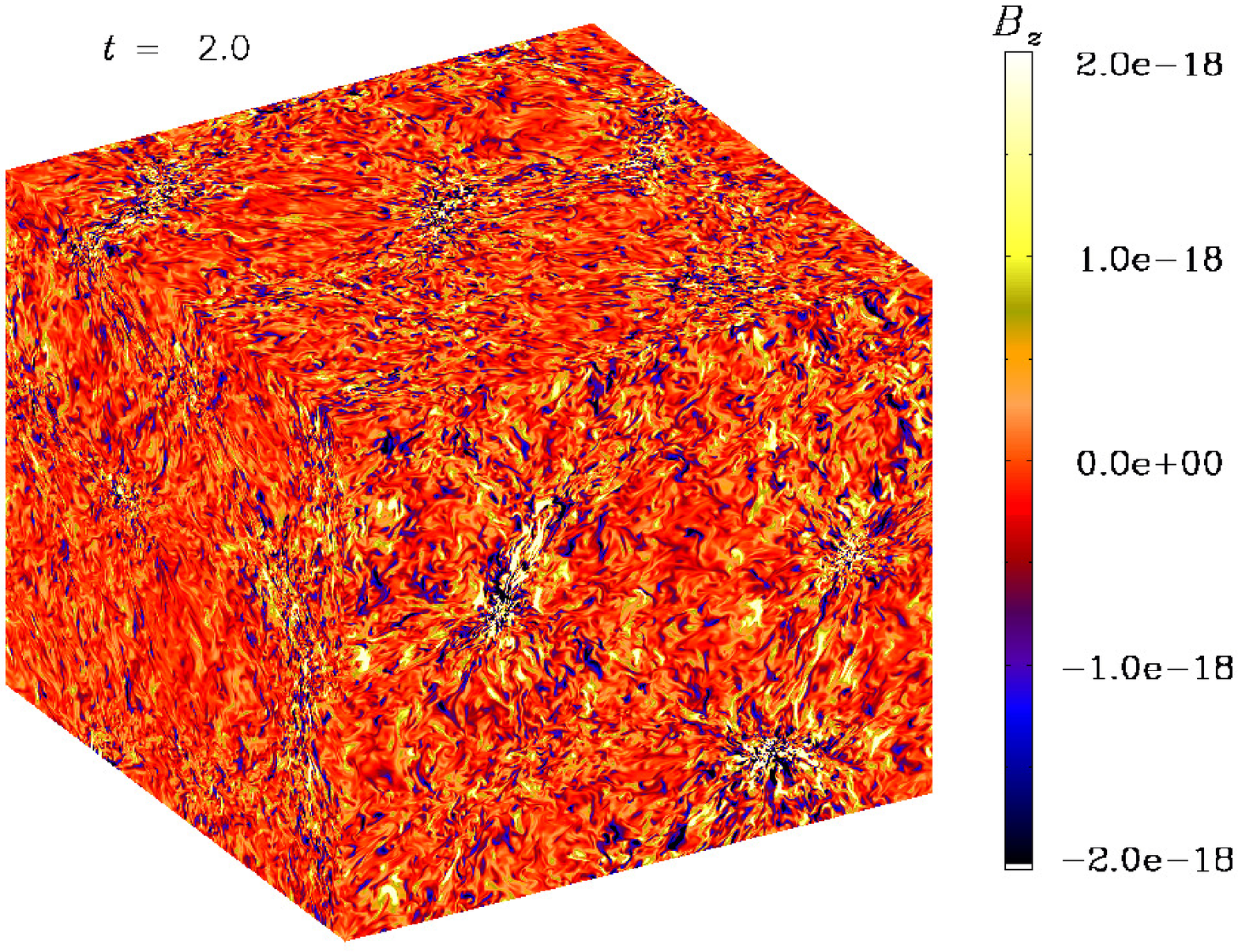}
\includegraphics[width=.49\textwidth]{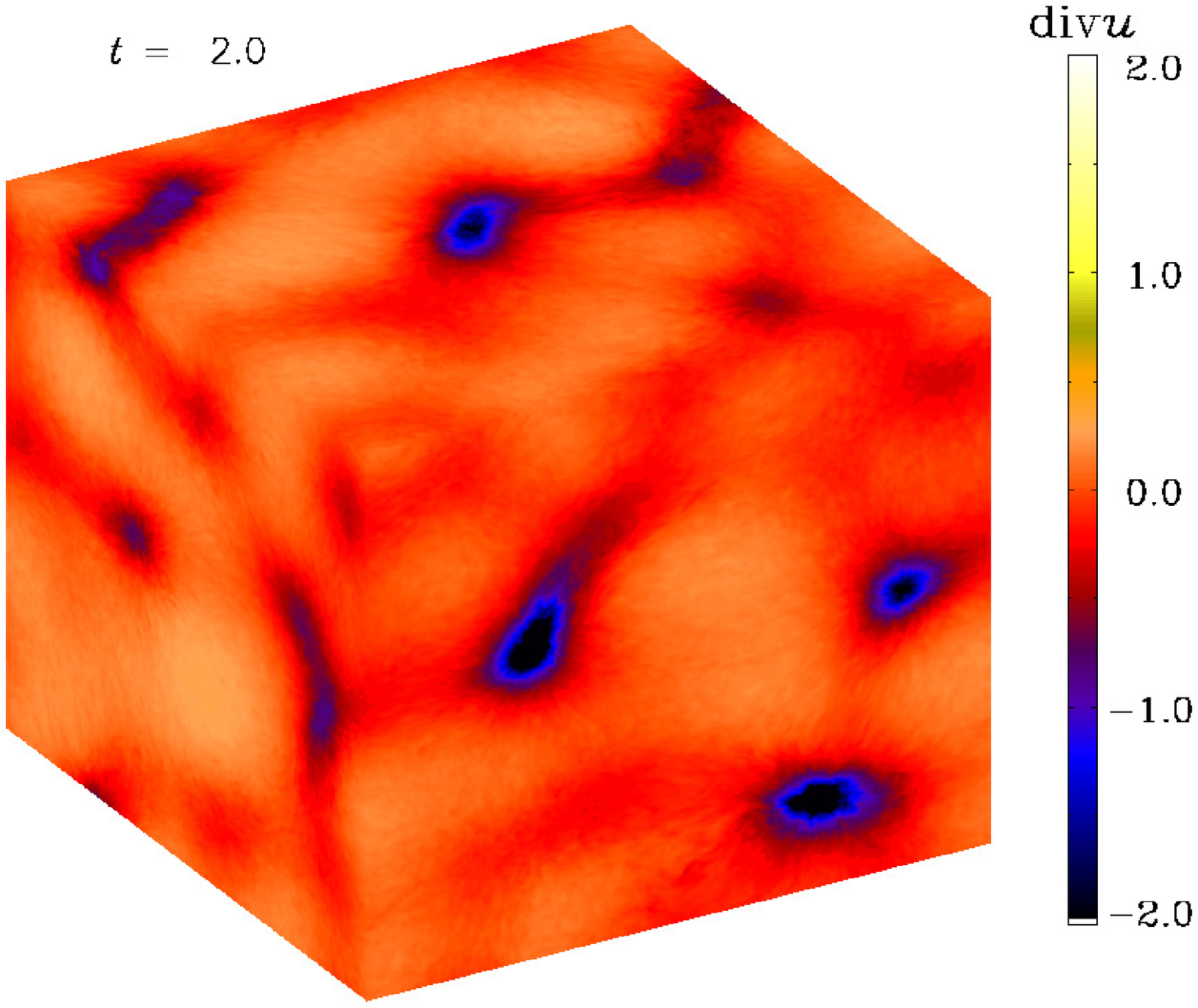}
\includegraphics[width=.49\textwidth]{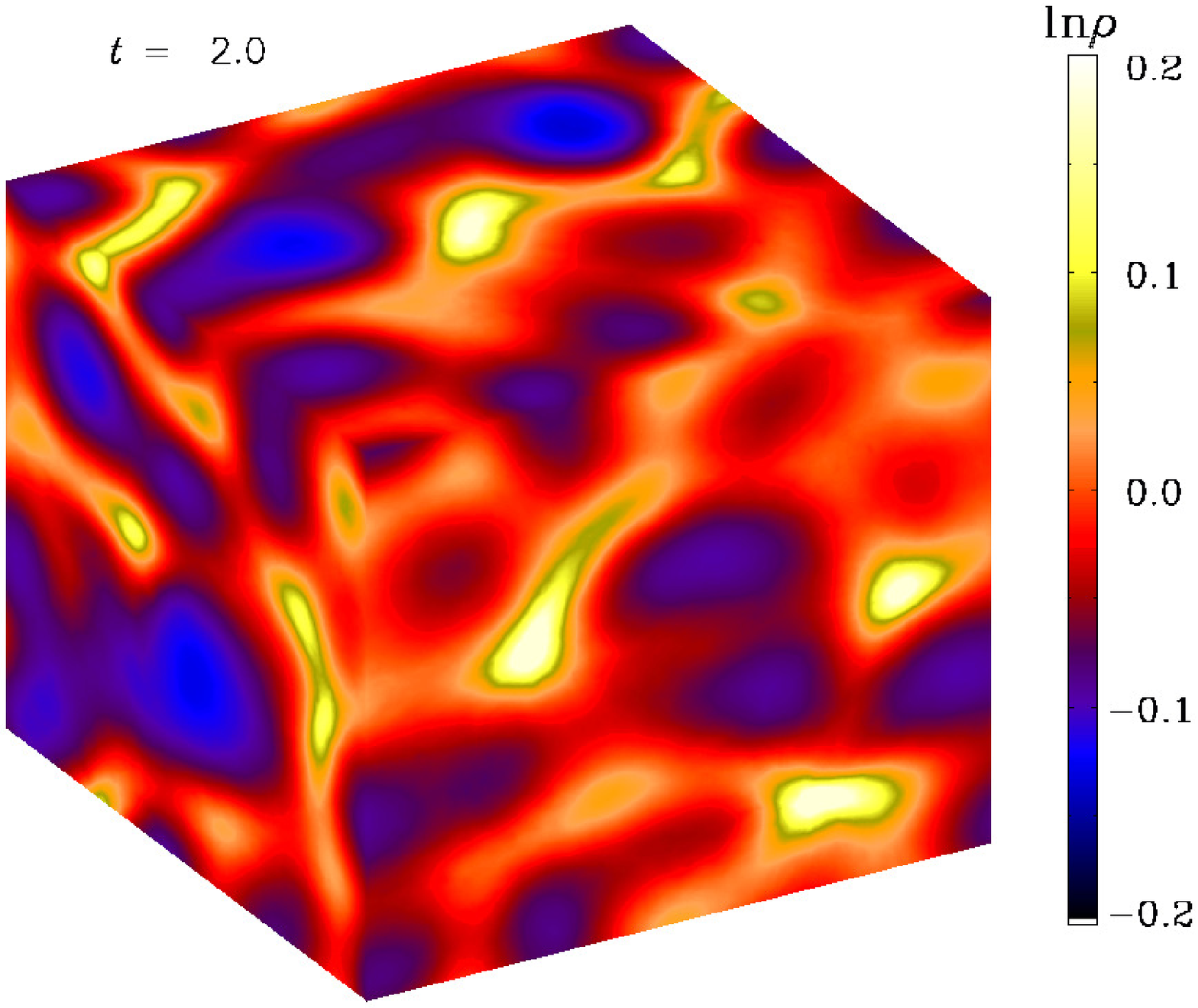}
\end{center}\caption[]{
$\omega_z$ (upper left), $B_z$ (upper right), $\nab\cdot\uu$ (lower left),
and $\ln\rho$ (lower right) near the end of the run.
Note the close correlation of the magnetic field with the vorticity
and their concentration toward regions of strong flow convergence
($\dive\uu<0$) and high density.
}\label{oo3_0020}\end{figure*}

\subsection{Numerical simulations}

We use the {\sc Pencil Code} \citep{JOSS}, which employs sixth order
accurate derivatives in space and a third order accurate time stepping
scheme.
Selfgravity was implemented by \cite{Joh+07} for modeling planetesimal
formation and employs Fourier transformation.
That same module is also being used for studying dust formation in the
ISM \citep{Mattsson+Hedvall22}.

\begin{figure*}\begin{center}
\includegraphics[width=\textwidth]{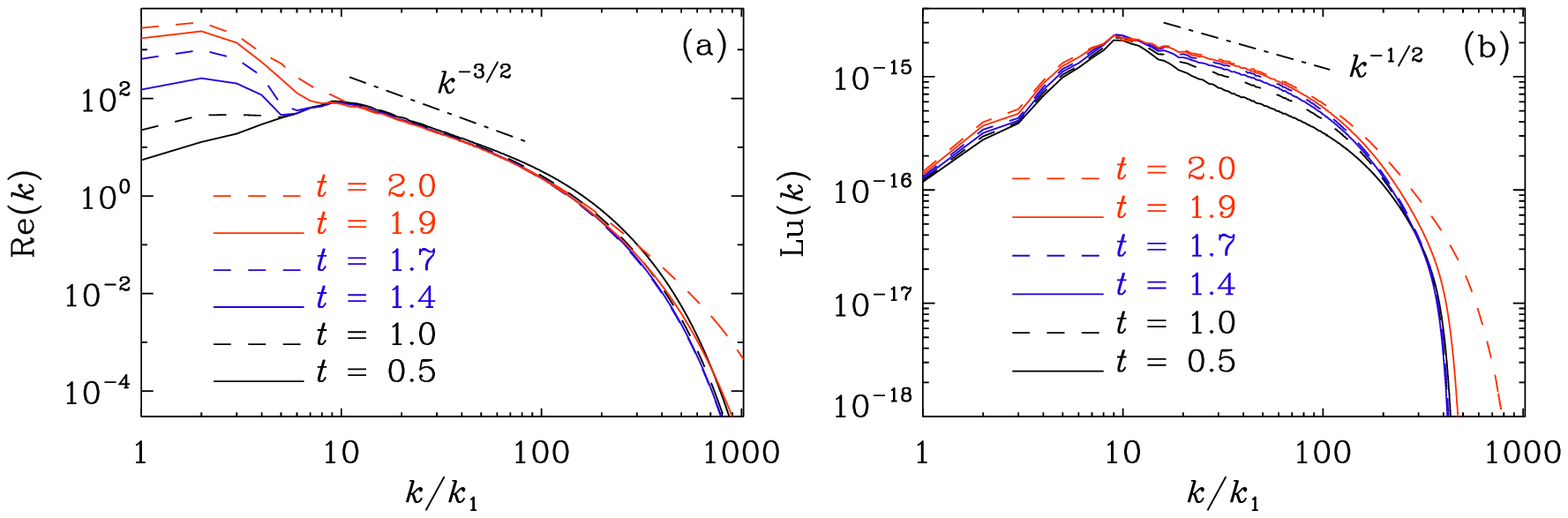}
\end{center}\caption[]{
(a) $\Rey(k)$ and (b) $\Lu(k)$ for Run~B at six different times,
indicated by line types and color.
}\label{rspec_select_S2048h_25_kf10e}\end{figure*}

\begin{table*}\caption{
Energy flux ratios and Reynolds numbers for the runs discussed in the paper.
Here, $k_{\rm J}$ and $\kf$ are in units of $k_1$.
}\vspace{12pt}\centerline{\begin{tabular}{ccccccccccccrcc}
Run & $k_{\rm J}$ & $\kf$ & $\Rey_{\kf}$ & $\Rey_{t_*}$ & $\Lu_{t_*}$ &
$\epsilon_{\rm J}^{\rm P}$ &
$\epsilon_{\rm J}^{\rm+K}$ &
$\epsilon_{\rm J}^{\rm-K}$ &
$\epsilon_{\rm J}^{\rm L}$ &
$\epsilon_{\rm L}^{\rm+M}$ &
$\epsilon_{\rm J}^{\rm-M}$ &
$\epsilon_{\rm M}^\Delta$ &
$\gamma$ &
$N^3$ \\
\hline
O1&2&10 & 500 &1000 &$2.3\times10^{-13}$& 0.30 & 0.68 & 0.02 & 0.000 &~~0.28  & 0.72 & 0.40 & $~~1.39$ & $2048^3$ \\
O2&5& 2 & 100 &1000 &$3.9\times10^{-14}$& 0.32 & 0.67 & 0.01 & 0.000 &~~0.49  & 0.51 & 0.96 & $~~0.83$ & $1024^3$ \\
A &5&10 & 500 &1000 &$1.0\times10^{-13}$& 0.33 & 0.64 & 0.03 & 0.000 &~~0.35  & 0.65 & 0.70 & $~~3.0$  & $2048^3$ \\
B &5&10 & 100 & 200 &$9.9\times10^{-15}$& 0.34 & 0.63 & 0.03 & 0.000 &~~0.21  & 0.79 & 0.26 & $~~0.44$ & $2048^3$ \\
b &5&10 & 100 & 200 &$9.9\times10^{-15}$& 0.31 & 0.66 & 0.03 & 0.000 &~~0.20  & 0.76 & 0.32 & $~~0.54$ & $1024^3$ \\
C &5&10 &  20 &  40 &$9.6\times10^{-16}$& 0.34 & 0.63 & 0.03 & 0.000 &~~0.06  & 0.94 & 0.06 & $~~0.08$ & $2048^3$ \\
D &5&10 &   5 &  10 &$1.1\times10^{-16}$& 0.31 & 0.66 & 0.03 & 0.000 &$-0.82$~~&1.82~~&$-0.48$& $-0.44$& $1024^3$ \\
E &5&10 &   1 &   2 &$5.6\times10^{-18}$& 0.25 & 0.73 & 0.02 & 0.000 &$-11.5$~~&12.5~~&$-0.96$& $-1.13$& $1024^3$ \\
S &5&10 & 500 &1300 &$8.7\times10^{-14}$& 0.31 & 0.60 & 0.09 & 0.000 &~~0.34  & 0.66 & 0.79 & $~~2.47$ & $1024^3$ \\
\hline
M1&5&10 & 500 &1000 &$1.3\times10^{3}$  & 0.36 & 0.46 & 0.13 & 0.05 &$-17.6$ & 18.6  &$-0.94$& $-0.38$  & $2048^3$ \\
M2&5&10 & 500 &1000 &$6.4\times10^{2}$  & 0.31 & 0.67 & 0.01 & 0.01 &$-0.97$ & 1.97  &$-0.45$& $-0.16$   & $2048^3$ \\
M3&5&10 & 500 &1000 &$4.1\times10^{2}$  & 0.32 & 0.65 & 0.01 & 0.02 &~~0.03  & 0.97  & 0.17 & $~~0.11$  & $2048^3$ \\
M4&5&10 & 100 & 200 &$9.8\times10^{0}$  & 0.33 & 0.65 & 0.02 & 0.00 &~~0.20  &  0.80 & 0.25 & $~~0.42$ & $2048^3$ \\
I1&5&10 & 500 &1000 &$1.6\times10^{3}$  & 0.29 & 0.68 & 0.01 & 0.02 &~~0.67  &  0.33 & 2.63 & $~~0.05$ & $2048^3$ \\
I2&5&10 & 500 &1000 &$7.5\times10^{2}$  & 0.31 & 0.67 & 0.01 & 0.01 &~~0.20  &  0.80 & 0.26 & $~~0.03$ & $2048^3$ \\
I3&5&10 & 500 &1000 &$4.3\times10^{2}$  & 0.32 & 0.65 & 0.01 & 0.02 &~~0.32  &  0.68 & 0.60 & $~~0.33$ & $2048^3$ \\
\label{Tsummary}\end{tabular}}\end{table*}

Many of the diagnostic quantities are calculated during run time,
including spectra and slices.
Most of the secondary data that are used for the plots are publicly
available; see the code and data availability statement at the end of
the paper.

\section{Results for weak magnetic fields}
\label{Results}

\subsection{Visualizations and spectra}
\label{Visualizations}

In \Fig{pcomp_urms2} we show the evolution of the Mach number and the rms
Alfv\'en speed normalized to the sound speed for $\sigma_{\rm J}=2\cs k_1$
(Run~O1) and $5\cs k_1$ (Runs~A--E), so $t_{\rm ff}\cs k_1=0.96$
and $0.38$, respectively.
In both cases, an exponential growth of $\Ma$ commences at some time.
For Run~O1, the growth rate agrees with that expected from the dispersion
relation, i.e., $\sigma/\cs k_1=\sqrt{3}$, but for Run~B, the actual value is
10\% smaller than the theoretically expected value, $\sigma/\cs k_1=\sqrt{24}$,
which could be related to the finite viscosity.
We define the moment when the rms velocity has recovered to its initial
value (denoted by the horizontal line for $\urms(0)/\cs=0.2$) as $t_*$.
Those characteristic times ($t_*\cs k_1\approx1.5$ and 4.9 for
$\sigma_{\rm J}/\cs k_1=5$ and 2, respectively) are denoted
by vertical dashed-dotted lines in the corresponding colors.
Those times correspond approximately to the moment when the negative
potential energy density begins to exceed the kinetic energy density,
i.e., when the virial parameter $\alpha_{\rm vir}$ drops below two;
see \App{Virial} for a demonstration.
In the supersonic case of Run~S, the collapse is found to occur earlier.
This is mainly a consequence of the stronger initial perturbations \citep[see, e.g.][for a review]{MacLow_Klessen2004}.

The growth of the magnetic field is quantified by the ratio $\vA/\cs$;
see \App{Growth} for a comparison with other measures such as $\Brms$ and
the ratio $|\BB|/\rho^{2/3}$.
Given that the velocity is decaying during the first part of the
evolution, we cannot expect an exponential growth of the magnetic field.
During the second part, when the velocity is exponentially increasing,
the magnetic field does not show a corresponding increase.
In the following, we tentatively associate the slow growth of the magnetic
field during the decay phase of the velocity field with a dynamo, and
the second part, which is dominated by the Jeans instability, with just
compressional amplification.
With these observations in mind, we continue using the term dynamo, but leave
it for further analysis to establish more rigorous and convincing criteria.

\begin{figure*}\begin{center}
\includegraphics[width=.49\textwidth]{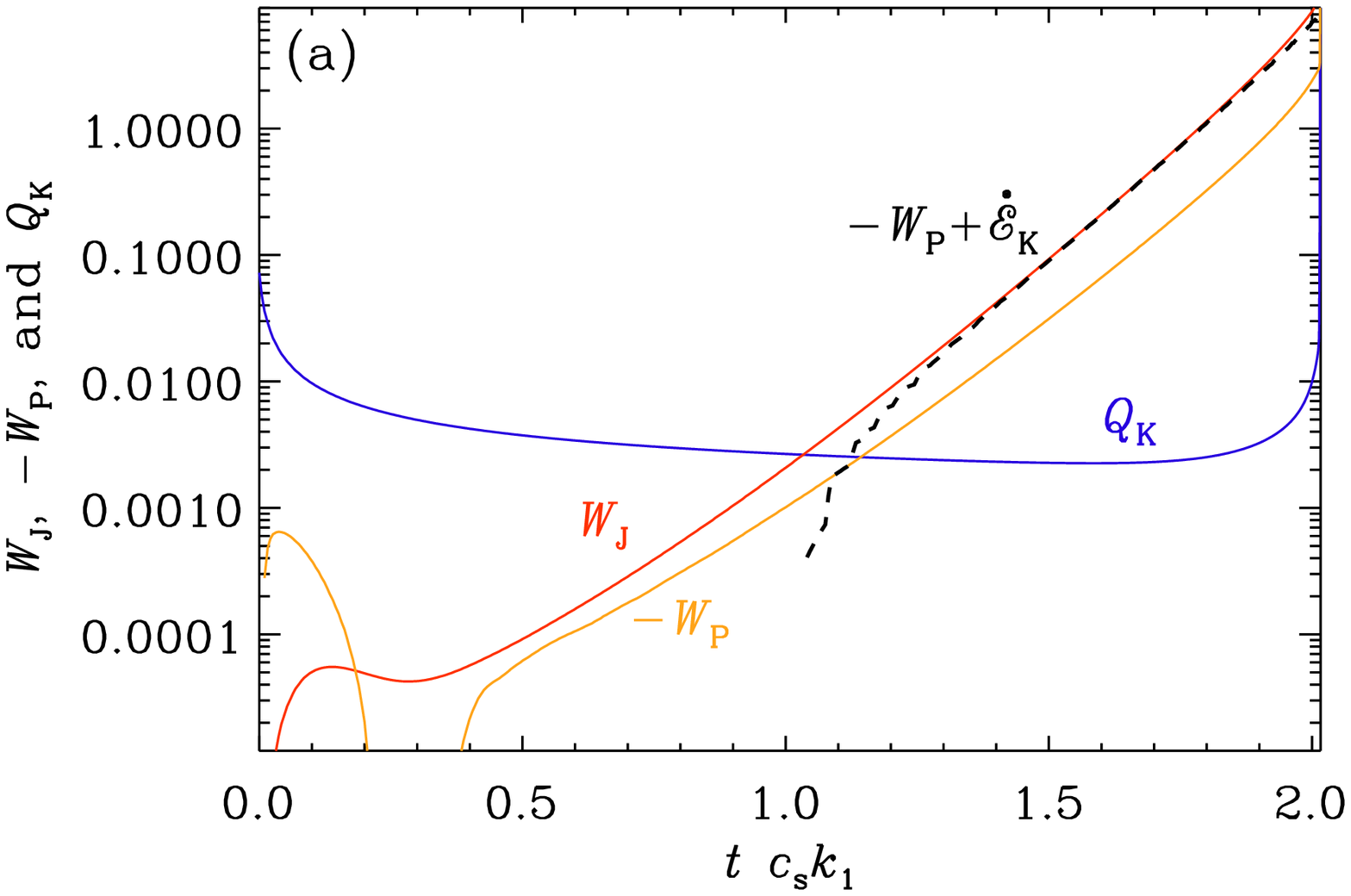}
\includegraphics[width=.49\textwidth]{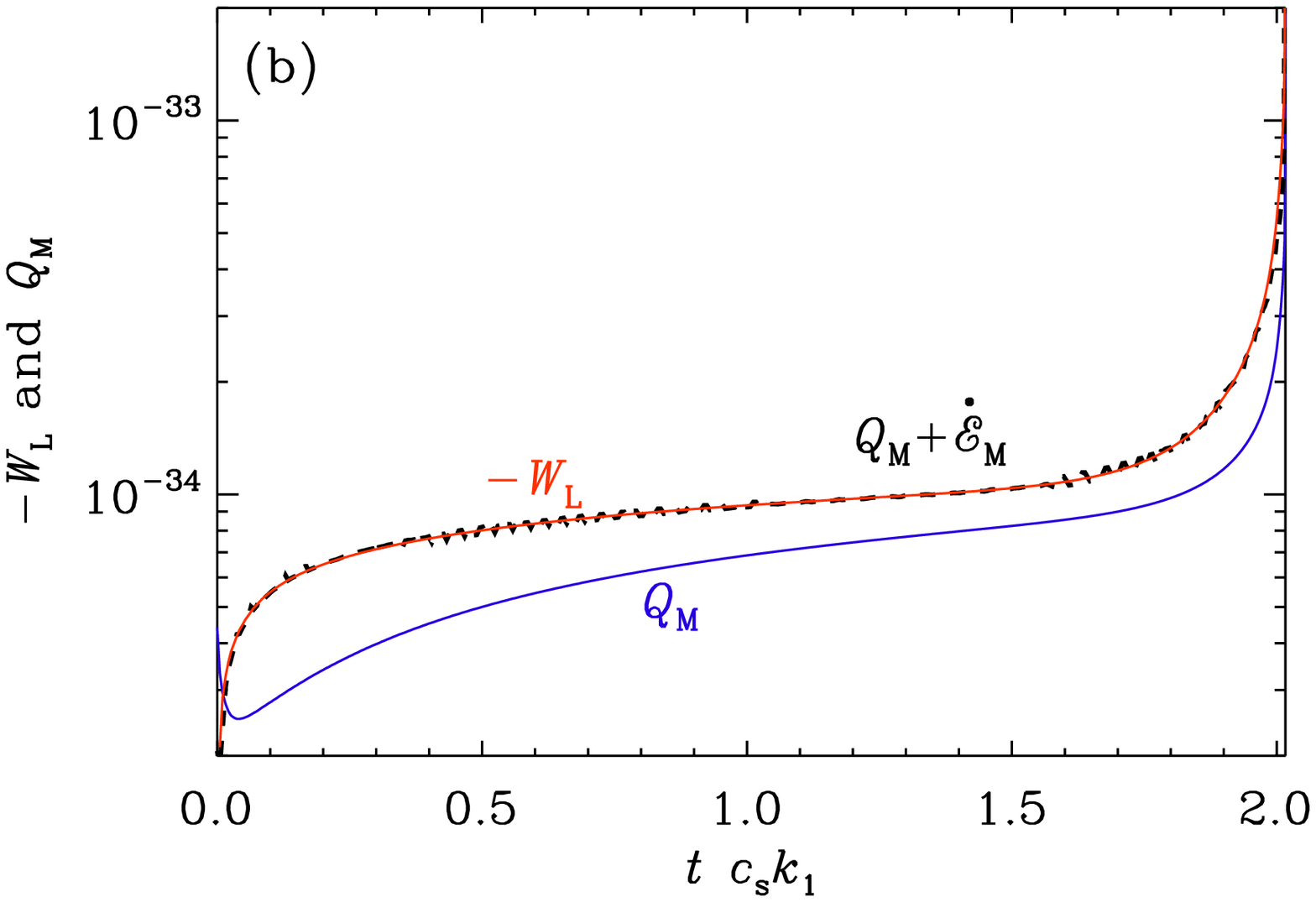}
\end{center}\caption[]{
(a) Work terms $W_{\rm J}$ (red) and $-W_{\rm P}$ (orange),
as well as $-W_{\rm P}+\dot{\cal E}_{\rm K}$ (black dashed).
Also shown is the kinetic energy dissipation, $Q_{\rm K}$.
(b) Work terms $-W_{\rm L}$ (red) and $Q_{\rm M}$ (blue),
as well as $Q_{\rm M}+\dot{\cal E}_{\rm M}$ (black dashed).
We recall that all work terms are in units of $\rho_0\cs^3 k_1$.
}\label{pdiss_S2048h_25_kf10e}\end{figure*}

For the rest of the paper, we focus on the case $\sigma_{\rm J}=5\cs k_1$,
so we expect growth for $k<k_{\rm J}\equiv5k_1$.
We summarize our runs in \Tab{Tsummary}.
In \Fig{oo3_0020} we show a visualization of the $z$ components of the
vorticity and magnetic field for Run~B, as well as its flow divergence and the
logarithmic density near the end of the run at $t\cs k_1=2.02=5.24\,t_{\rm ff}$,
shortly before further compression can no longer be resolved.
The fact that this time is much longer than unity is due to
the periodicity of the solution to the Poisson equation
\citep{Federrath+12,Lane+21}.
We see that $\omega_z$ and $B_z$ show strong concentrations toward
regions where the density also increases.
Note that the regions of negative flow divergence are more
strongly concentrated that those of positive flow divergence, but both
$\nab\cdot\uu$ and $\ln\rho$ are dominated by a spatially smooth component
that, unlike $\omega_z$ and $B_z$, lack small scales.
Very weak small-scale perturbations can be seen in the visualizations
of $\nab\cdot\uu$, but not in those of $\ln\rho$.
Gradients of $\ln\rho$ do, however, show marked small-scale structure.
In our runs, the density contours are more often aligned with the magnetic
field vector than being perpendicular to it.
This is demonstrated in \App{Growth} for Runs~M1 and I1, where we see
that $\BB$ is mostly perpendicular to $\nab\ln\rho$.

To see whether at any scale, the Reynolds number is large enough
for dynamo action, we employ the $k$-dependent Reynolds numbers.
In \Fig{rspec_select_S2048h_25_kf10e}, we show spectra of $\Rey(k)$
and $\Lu(k)$ at different times.
We clearly see that the spectra display instability for
$k<k_{\rm J}=5\,k_1$, and only at late times ($t\cs k_1>1.7$),
somewhat smaller scales begin to grow as well.
However, the effect on the rest of the spectrum is surprisingly weak.
There is a small increase of $\Lu(k)$ at all wavenumbers, but there is
no visible effect from the Jeans instability itself, except for the time
close to the end of the simulation where one sees an increase of both
$\Rey(k)$ and $\Lu(k)$ at the highest wavenumbers, indicating that more
energy is now being channeled through the turbulent cascade.
We also see that the value of $\Rey(k)$ near the wavenumber $\kf$, where
the initial kinetic energy spectrum peaks, is around $\Rey_{\kf}=100$.
According to previous studies, this value is high enough for dynamo action.
Our more detailed studies below confirm that this is indeed the case.
When changing $\nu$ and $\eta$, the peak values of $\Rey(k)$ and $\Lu(k)$
change correspondingly.
In \App{Relam}, we show that $\Rey_{\kf}=100$ is close (within a factor of two)
to the Taylor microscale Reynolds number.

\subsection{Work terms}
\label{WorkTerms}

Next, we consider the evolution of the various work terms; see
\Fig{pdiss_S2048h_25_kf10e}.
The work done by the gravity force, $-\bra{\rho\uu\cdot\nab\Phi}\equiv W_{\rm J}>0$,
leads to flow compression, $\bra{p\nab\cdot\uu}\equiv W_{\rm P}<0$,
and an increase of the kinetic energy density, so
\EQ
-\bra{\rho\uu\cdot\nab\Phi}=-\bra{p\nab\cdot\uu}+\frac{\dd\EEK}{\dd t}...,
\EN
where the ellipsis denotes the sum of two additional, subdominant terms:
$Q_{\rm K}$ and $-W_{\rm L}$.
The dominant balance in the kinetic energy evolution is given by the
Jeans work, which is found to be balanced to 1/3 by the pressure work
and to 2/3 by the growth in kinetic energy, i.e.,
\EQ
-W_{\rm P}\approx\frac{1}{3}W_{\rm J},\quad
\dot{\cal E}_{\rm K}\approx\frac{2}{3}W_{\rm J}.
\label{JeansRatios}
\EN
The latter can be integrated to give
${\cal E}_{\rm K}\approx(2/3)\int W_{\rm J}\,\dd t$.
Likewise, integrating \Eq{dEP} gives
$-{\cal E}_{\rm P}\approx\int W_{\rm J}\,\dd t$, which implies
$\alpha_{\rm vir}=2{\cal E}_{\rm K}/|{\cal E}_{\rm P}|\approx4/3$.
Its value would be unity, if only half of $W_{\rm J}$ went into the
growth of kinetic energy, but this is not the case.
It is important to realize that the energy flux ratios in
\Eq{JeansRatios} apply to the time $t=t_*$.
They change at later times toward 1/4 and 3/4 for the pressure work
and growth in kinetic energy; see \App{Virial}, which implies
$\alpha_{\rm vir}\approx3/2$.

These ratios of the work terms implies that about one-third of the
gravitational energy goes into compressional heating and two-thirds go
into kinetic energy before eventually also being thermalized.
At the reference time $t_*=1.5/(\cs k_1)$, however, viscous dissipation
contributes only about 3\%; see Runs~A--E in \Tab{Tsummary}.

\begin{figure}\begin{center}
\includegraphics[width=\columnwidth]{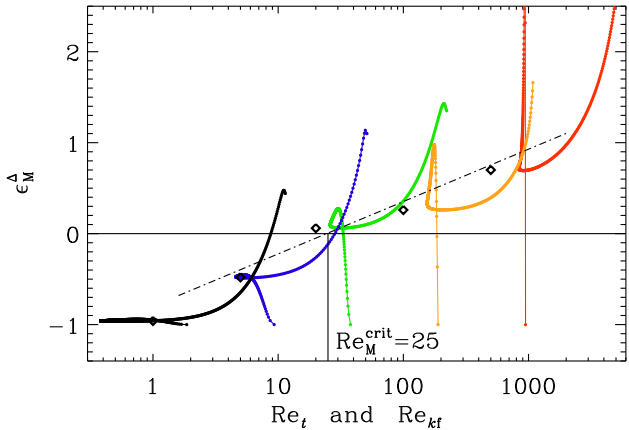}
\end{center}\caption[]{
Time dependence of $(-W_{\rm L}/Q_{\rm M})-1$ on $\Rey_t$
for Runs~A (red), B (orange), C (green), D (blue), and E (black).
The values of $(-W_{\rm L}/Q_{\rm M})-1$ from \Tab{Tsummary} are shown as
diamonds as a function of $\Rey_{\kf}$ and are seen to obey an approximate
fit given by $\ln(0.45\,\Rey^{1/4})$; see the dashed-dotted line.
}\label{pcomp_diss}\end{figure}

In the kinematic regime of the dynamo, the work done by the
Lorentz force is very small; see \Fig{pdiss_S2048h_25_kf10e}(b).
Nevertheless, this term exceeds the Joule dissipation when
$\Rey$ is large enough.
Here, the work done {\em against} the Lorentz force, $-W_{\rm L}$, leads
to Joule heating and an increase in the magnetic energy density, i.e.,
\EQ
-\bra{\uu\cdot(\JJ\times\BB)}=\bra{\mu_0\eta\JJ^2}+\frac{\dd\EEM}{\dd t}.
\EN
Therefore, based on the positivity of $(-W_{\rm L})-Q_{\rm M}$, i.e.,
the positivity of $\gamma$, we argue that we can determine the threshold
for dynamo action to be around 25.
This is the value for which the interpolated line of
$(-W_{\rm L}/Q_{\rm M})-1$ vs $\Rey$ crosses zero,
which corresponds to marginal dynamo excitation; \Fig{pcomp_diss}.
In fact, Run~C with $\Rey_{\kf}=20$ is close to the marginal point;
see \Tab{Tsummary}.

During the exponential growth phase of the Jeans instability,
the velocity grows at the rate $\sqrt{24}\cs k_1$.
During that period, $\Rey(t)$ increases rapidly with time, and
so does also the difference $(-W_{\rm L})-Q_{\rm M}$.

\begin{figure}\begin{center}
\includegraphics[width=\columnwidth]{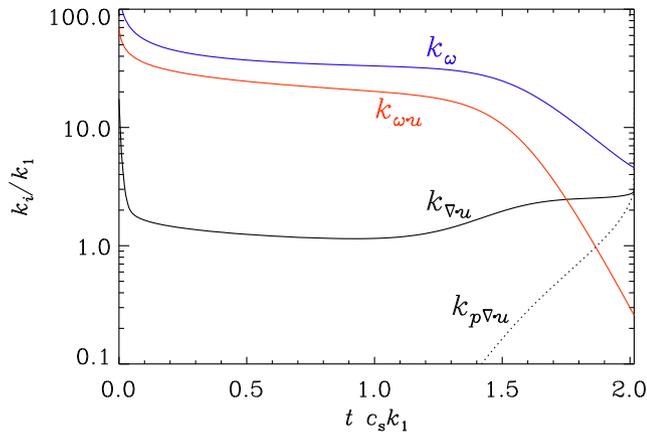}
\end{center}\caption[]{
Characteristic wavenumbers $k_\omega$ (blue),
$k_{\oo\cdot\uu}$ (red), $k_{\,\nab\cdot\uu}$ (solid black),
and $k_{p\nab\cdot\uu}$ (dotted black).
}\label{pvort_S2048h_25_kf10e}\end{figure}

\begin{figure*}\begin{center}
\includegraphics[width=.49\textwidth]{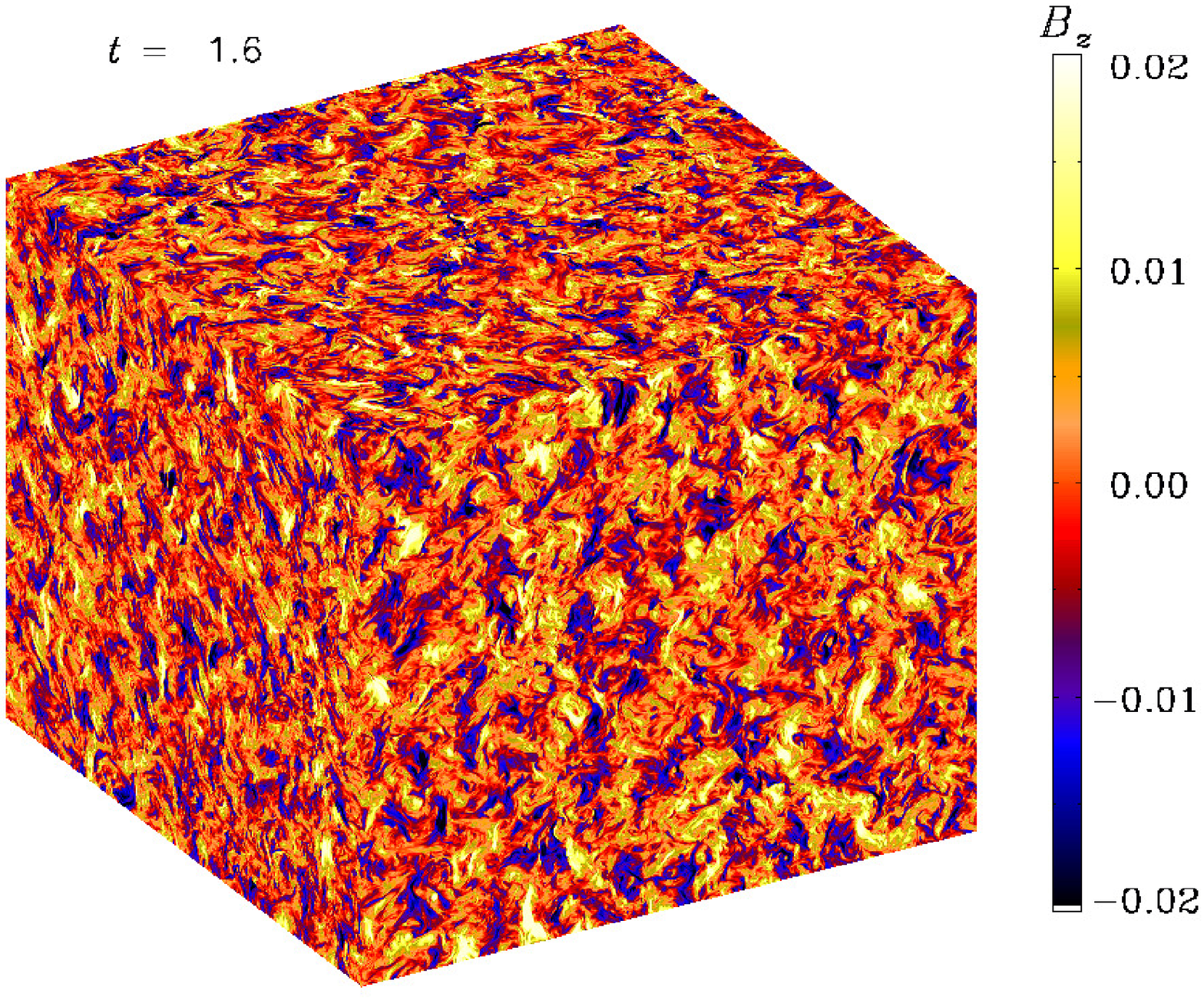}
\includegraphics[width=.49\textwidth]{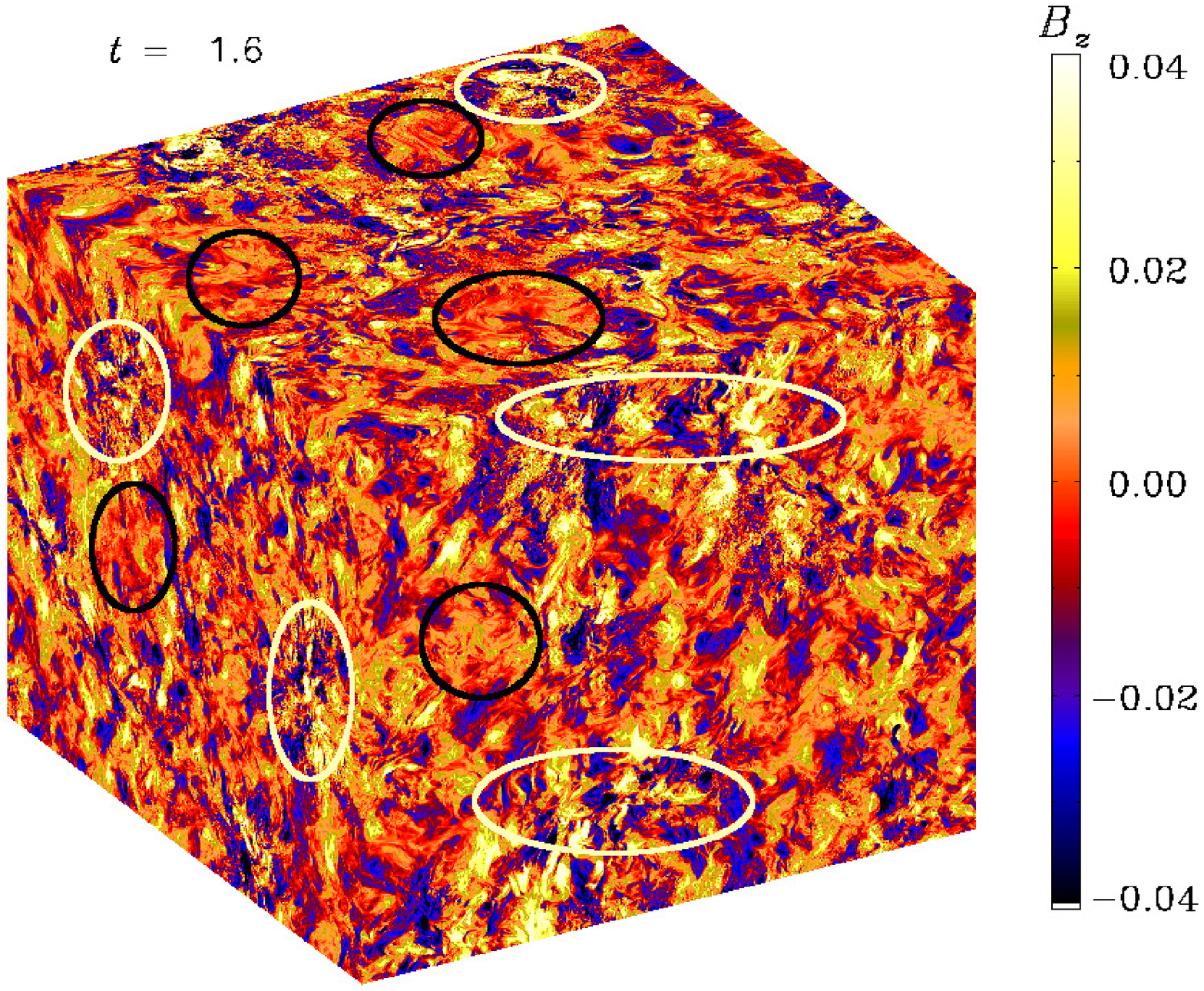}
\end{center}\caption[]{
$B_z$ for Runs~M2 (left) and M1 (right) at $t\cs k_1=1.6$.
Note the large-scale concentrations (marked by white ellipses)
and voids (marked by black circles) for stronger fields,
even though the stage of the collapse is the same.
For Run~M1, one sees indications that the $2048^3$ resolution
begins to become insufficient at $t=1.6$.
}\label{bb3_0016}\end{figure*}

\subsection{Dependence on the magnetic Reynolds number}
\label{DepReynolds}

As the magnetic Reynolds number increases, we expect a dynamo to
become stronger and thus,
$\epsilon_{\rm M}^\Delta\equiv(-W_{\rm L}-Q_{\rm M})/Q_{\rm M}$ to increase.
However, the magnetic Reynolds number is time-dependent, because $\urms$
increases.
This raises the question whether this dependence follows
a similar trend that is seen by comparing different runs.

In \Fig{pcomp_diss} we plot $\epsilon_{\rm M}^\Delta(t)$
versus $\Rey_t$ and compare with the values listed in \Tab{Tsummary}
for the nominal time $t_*=1.5/(\cs k_1)$ versus $\Rey_{\kf}$.
We see that $\epsilon_{\rm M}^\Delta(t_*)$ shows a rather shallow increase
with $\Rm(t_*)$ of the form
\EQ
\epsilon_{\rm M}^\Delta\approx\ln(0.45\,\Rey^{1/4}).
\EN
The time-dependent tracks do approximately match this dependence at
around intermediate times, but all the lines are curved and shallower
for early times and steeper at late times, where the Jeans instability
becomes dominant.

\subsection{Compression and vorticity}

The magnetic field growth seems to be strongly correlated with the
rms vorticity.
Earlier studies showed that it is only the vortical part of the
flow that leads to dynamo action \citep{MB06}.
To quantify the relative importance of vortical and irrotational
or compressive contributions to the velocity field, we show in
\Fig{pvort_S2048h_25_kf10e} the evolution of the characteristic
wavenumbers $k_\omega$, $k_{\oo\cdot\uu}$, $k_{\,\nab\cdot\uu}$,
and $k_{p\nab\cdot\uu}$.

We see that $k_\omega\approx35$ and $k_{\oo\cdot\uu}\approx20$
during the early phase when the collapse velocity is still subdominant.
When the collapse becomes dominant, $k_\omega$ and $k_{\oo\cdot\uu}$
rapidly decline and $k_{\,\nab\cdot\uu}\approx2.5$ prior to the final
collapse.
Nevertheless, we always find $k_\omega>k_{\,\nab\cdot\uu}$, i.e.,
vorticity is still important.
This is partially because of the compression of vortices, as seen in
\Fig{oo3_0020}, which enhances the vorticity.
Furthermore, we find that $k_{p\nab\cdot\uu}\ll k_{\,\nab\cdot\uu}$,
except very close to the collapse when the two are similar.

\subsection{Dependence on resolution}

For our reference models, we 
use a fairly high resolution of $2048^3$ mesh points.
Larger resolutions become easily impracticable and reach computational
memory limitations.
However, the dependence on resolution is small and the maximum run time
before the collapse stops the calculation hardly changes at all when
increasing the resolution by a factor of two.
By comparing Runs~b and B, we see that the values of
$W_{\rm P}/W_{\rm J}$ and $\dot{\cal E}_{\rm K}/W_{\rm J}$,
as well as those of $\dot{\cal E}_{\rm M}/(-W_{\rm L})$ and
$Q_{\rm M}/(-W_{\rm L})$ agree with each other within 10\%.

\begin{figure*}\begin{center}
\includegraphics[width=.49\textwidth]{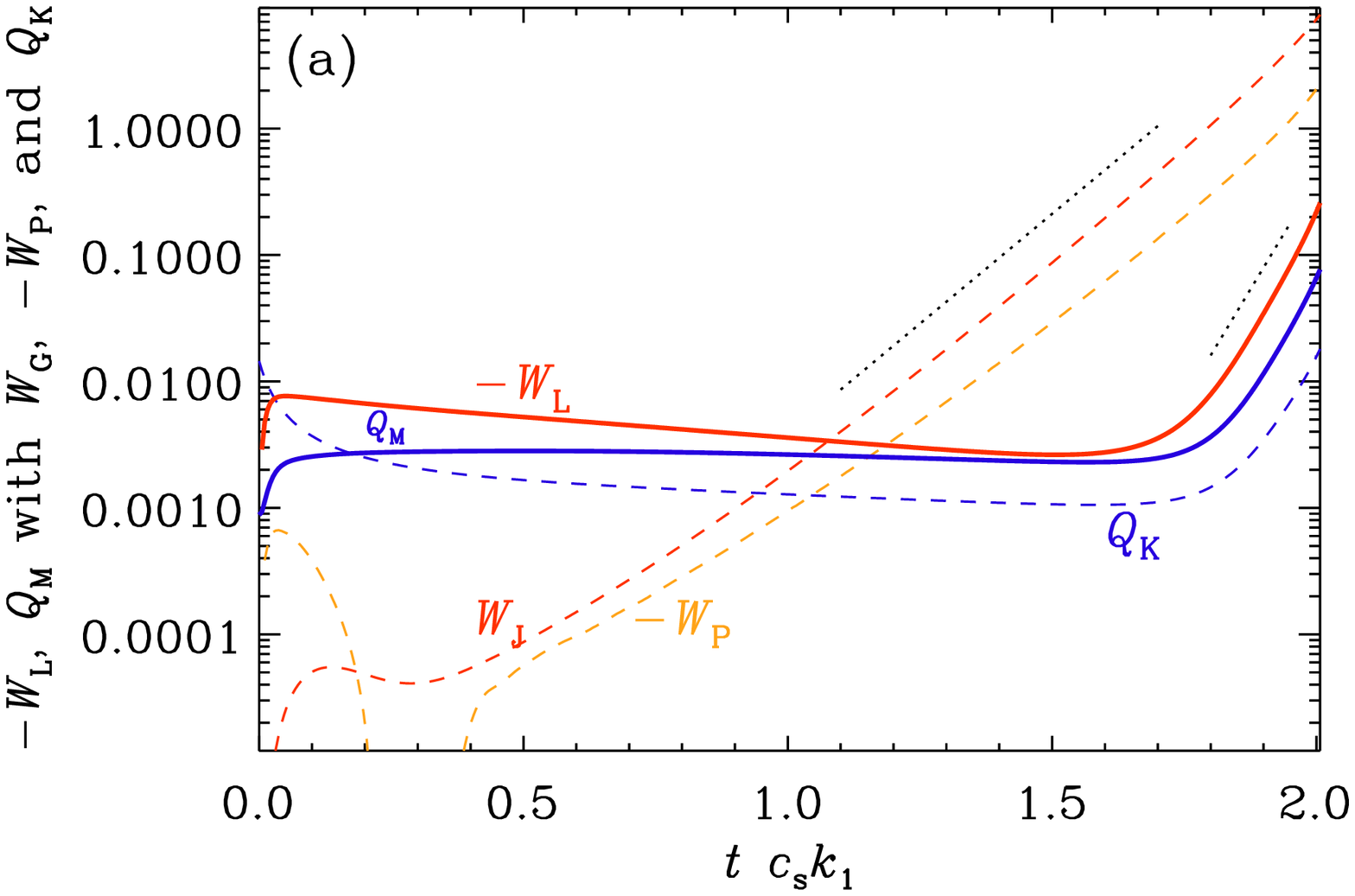}
\includegraphics[width=.49\textwidth]{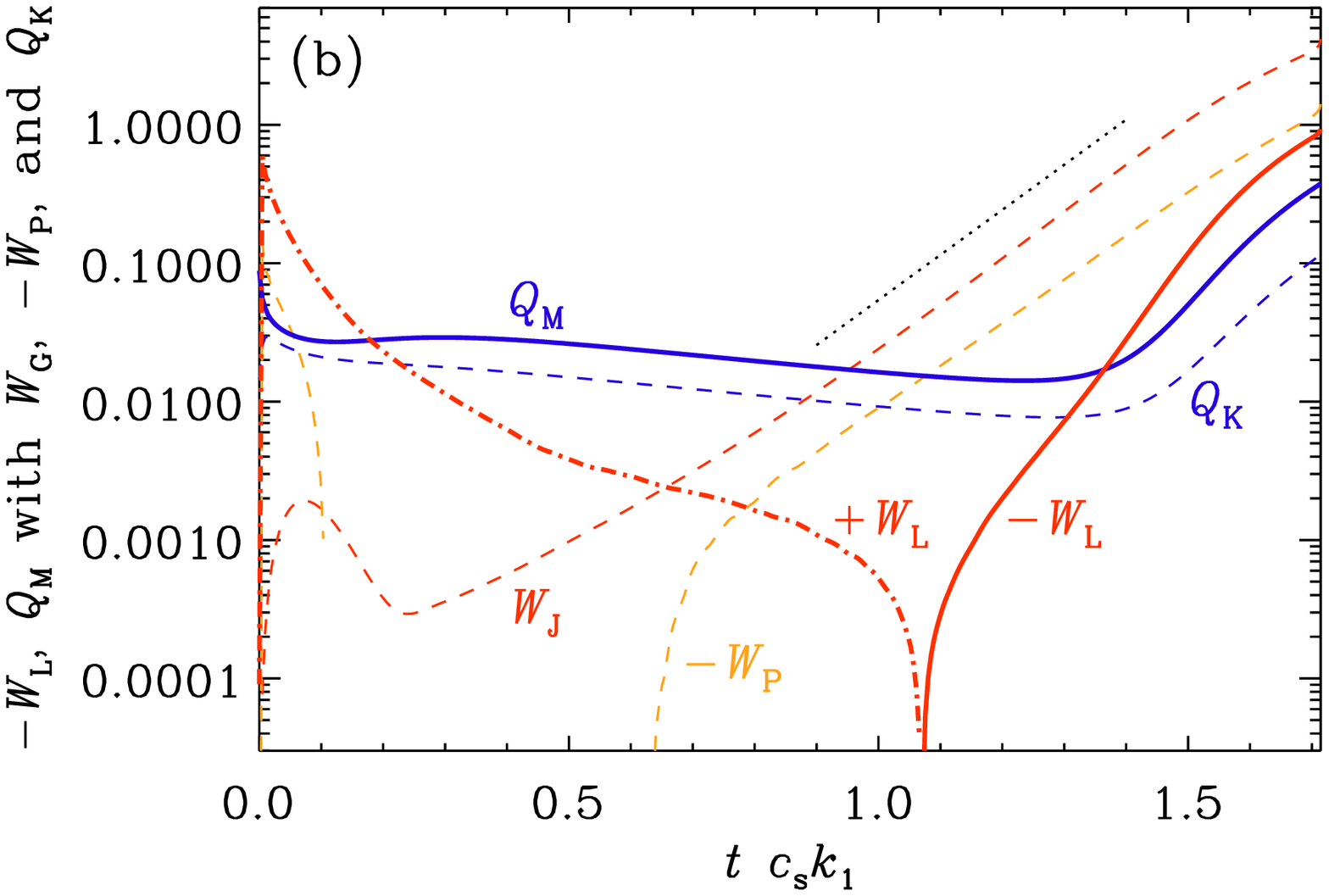}
\end{center}\caption[]{
Work terms $-W_{\rm L}$ (solid red) and $Q_{\rm M}$ (solid blue), together
with $W_{\rm J}$ (dashed red) and $-W_{\rm P}$ (dashed orange), as well as
$Q_{\rm K}$ (dashed blue) for (a) Run~M3 with a relatively weak magnetic field
and (b) Run~M1 with a strong magnetic field.
Dotted lines denote the slopes corresponding to the growth rates
$1.6\,\sigma_{\rm J}$ and $3.2\,\sigma_{\rm J}$ in panel (a) and
$1.5\,\sigma_{\rm J}$ in panel (b).
The thick dashed-dotted line in panel (b) denotes $W_{\rm L}>0$.
Note that the work terms are in units of $\rho_0\cs^3 k_1$.
}\label{pdiss_mag_S2048B_25_kf10f}\end{figure*}

\section{Results for strong magnetic fields}
\label{ResultsStrong}

\subsection{Earlier collapse with stronger fields}

We now consider cases where the magnetic field is dynamically important.
This situation is of particular interest for dense pre-stellar
cores, where the measured \citep[e.g.,][]{Crutcher2012} or inferred
\citep[e.g.,][]{Karoly2020,Pattle2021} magnetic fields are particularly
strong.

\Fig{bb3_0016} shows a comparison of the magnetic field patterns for
Runs~M3 and M1, i.e., for weaker and stronger fields, respectively.
For the stronger magnetic fields in Run~M1, the magnetic eddies appear
to be organized in larger patches that correspond to over- or underdense
regions.
For the stronger magnetic fields in Run~M1, the magnetic eddies appear
earlier than in Run~M3, which is probably the result of the magnetically
driven motion early on; see \Fig{pcomp_urms2}.
This behavior is suggestive of an accelerated collapse process.
This is an important difference to the standard paradigm of magnetically
controlled star formation that employs a uniform magnetic field
\citep{Mestel+Spitzer56,Mouschovias+Spitzer76,Shu77}.
Instead, here the magnetic field is turbulent and only has moderate
large-scale coherence.

\subsection{When $W_{\rm L}$ affects the collapse}

In \Figsp{pdiss_mag_S2048B_25_kf10f}{a}{b} we show the work terms for
Runs~M3 and M1, respectively.
Again, we see that a turbulent magnetic field does not systematically
delay the collapse, and a strong field can even accelerate it.

\begin{figure*}\begin{center}
\includegraphics[width=.49\textwidth]{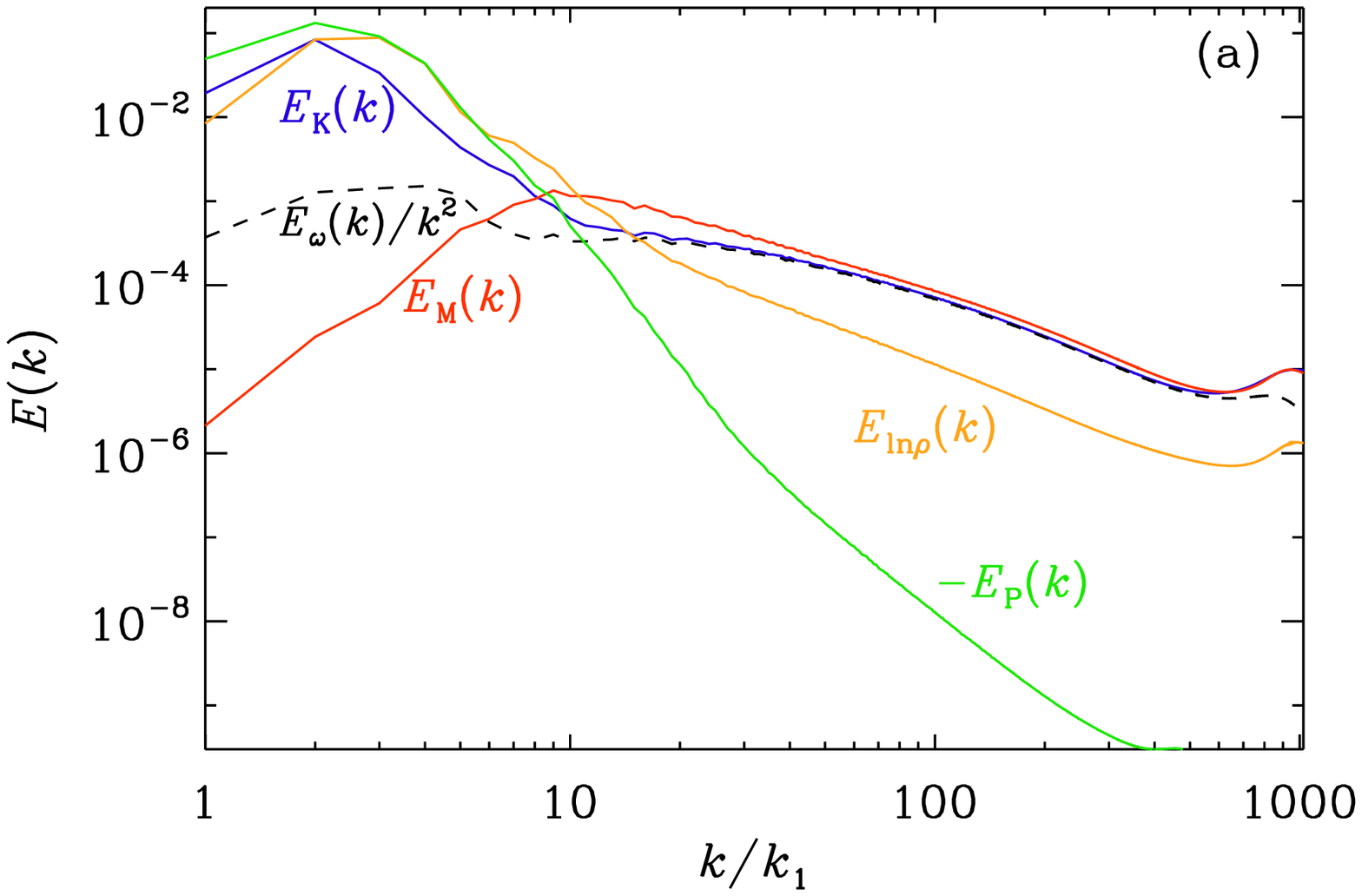}
\includegraphics[width=.49\textwidth]{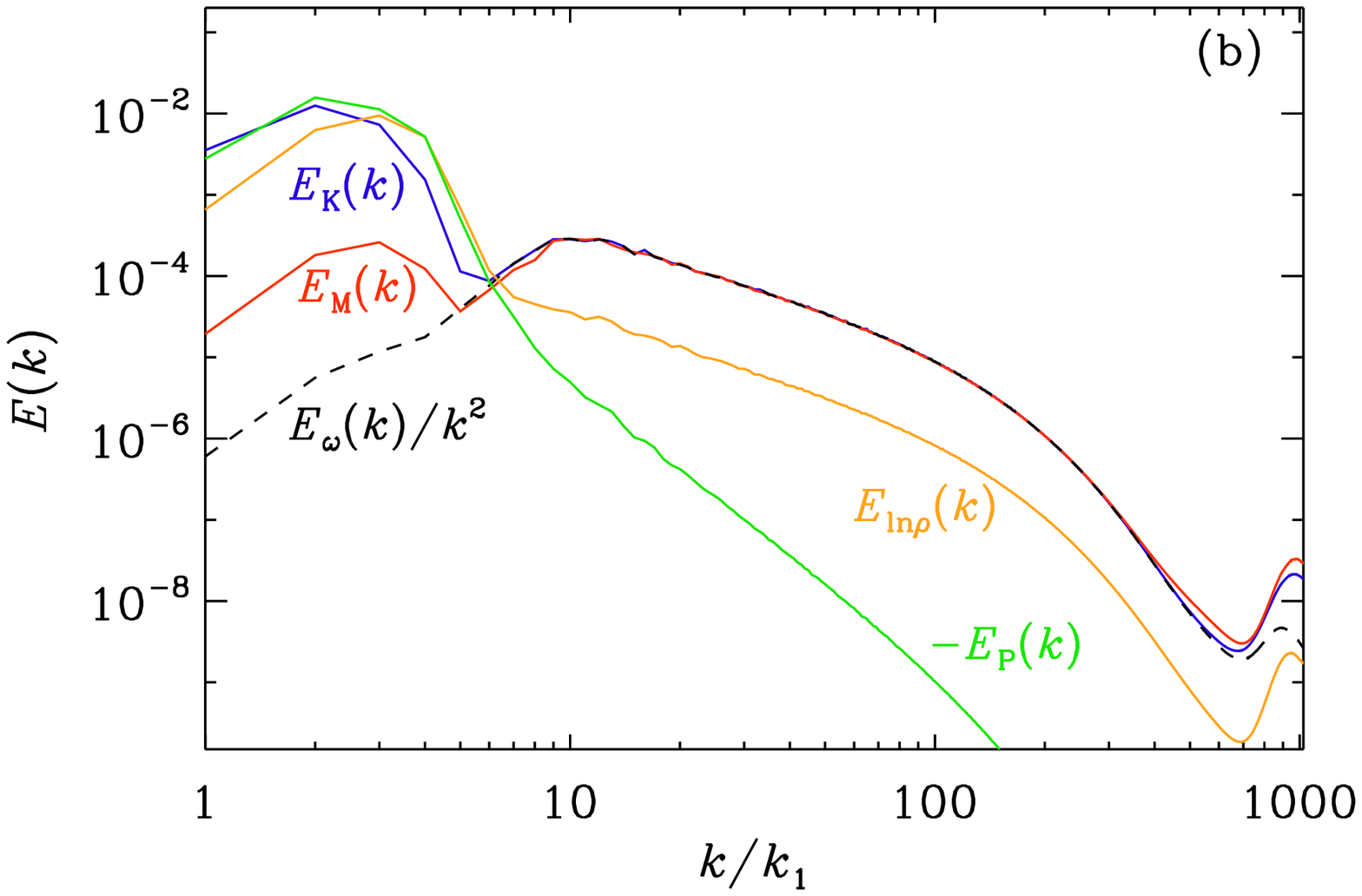}
\end{center}\caption[]{
Comparison of kinetic, magnetic, and potential energy and density spectra
at $t\cs k_1=1.7$ for (a) Run~M1 with a strong turbulent initial magnetic
field and (b) Run~I1 with a strong uniform initial magnetic field.
The dashed lines denoted with $E_{\oo}(k)/k^2$ correspond to kinetic
energy spectra of the vortical part of the velocity.
Here, the $E(k)$ are in units of $\cs^2 k_1^{-1}$.
Note that both axis ranges in (a) and (b) are the same.
}\label{rspec_last}\end{figure*}

We also see that the late-time exponential increase of $-W_{\rm L}$
and $Q_{\rm M}$ changes with respect to the weak-field behavior from
being twice the rate of $W_{\rm J}$ to being equal to it; see
\Figp{pdiss_mag_S2048B_25_kf10f}{b}.
Assuming the change in $\JJ\times\BB$ to be itself proportional to the
change in $\uu$, we see that $-W_{\rm L}$ is quadratic in $\uu$, which
explains the growth of $-W_{\rm L}$ at twice to Jeans rate.
Once the field is strong, $\JJ\times\BB$ does not change much anymore,
and so $-W_{\rm L}$ only grows at the Jeans rate.

The gravitational collapse is primarily characterized by the increase
in the velocity $\uu$ and not so much by a change in gravity.
In reality, however, the change in gravity does contribute about
50\%, and so we find that $W_{\rm J}$ increases at a rate of about
$1.6\,\sigma_{\rm J}$, and $-W_{\rm L}$ increases at a rate of about
$3.2\,\sigma_{\rm J}$.

\begin{figure*}\begin{center}
\includegraphics[width=\textwidth]{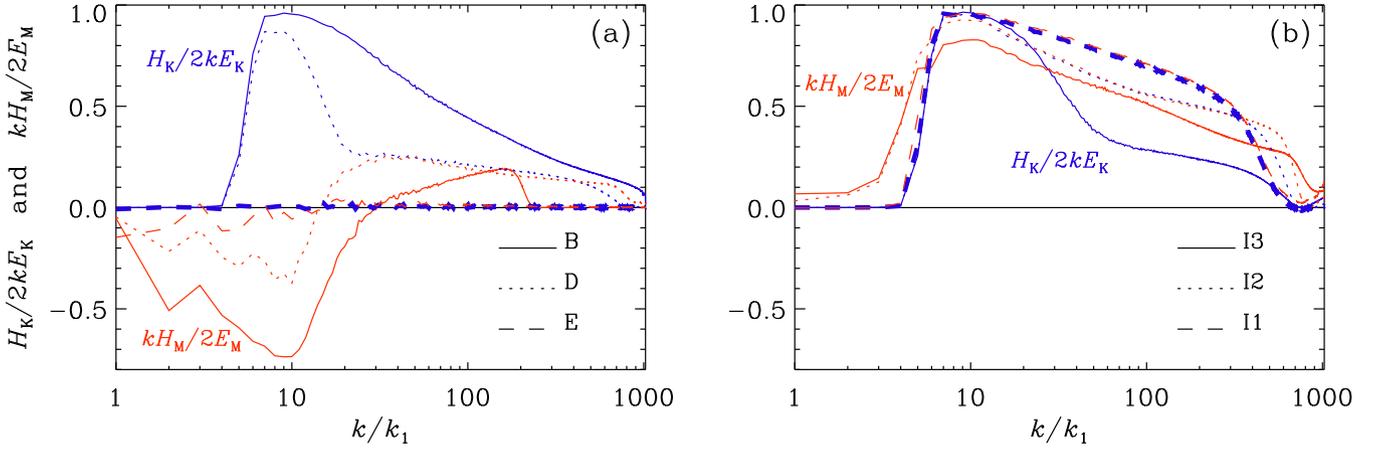}
\end{center}\caption[]{
Comparison of relative helicity spectra with (a) turbulent and (b) uniform
initial magnetic fields.
Blue (red) lines denote kinetic (magnetic) relative helicity spectra.
The solid lines are for Run~B (I3), the dotted lines are for Run~D (I2),
and the thick dashed lines are for Run~E (I1) for turbulent (uniform)
initial magnetic fields.
}\label{rspec_hel_comp}\end{figure*}

\subsection{Comparison with uniform initial magnetic fields}

In \Fig{rspec_last}(a) we show kinetic and magnetic energy spectra
together with logarithmic density spectra and the normalized enstrophy
spectra representing the kinetic energy spectra of the vortical part,
and compare with the case of a uniform initial magnetic field in
\Fig{rspec_last}(b).
The two panels correspond to Runs~M1 and I1.
Even though $\Lu_{t_*}$ is 25\% higher in I1,
the spectral energies are lower.
The density spectra show a rapid increase for $k\leq k_{\rm J}$, which
is associated with the Jeans instability, as was already seen in the
kinetic energy spectra; see \Figp{rspec_select_S2048h_25_kf10e}{a}.

The magnetic energy spectra and the kinetic energy spectra of the vortical
part do not show the same increase, but there is a slight one, which
is different for the cases with a turbulent and an initially uniform field.
To understand this, it is useful to discuss now the kinetic and magnetic
helicity spectra for the two cases.
We see that, for runs with an initially uniform magnetic field at
intermediate wavenumbers, the magnetic energy is either in nearly perfect
equipartition, or in super-equipartition with the kinetic energy.
However, it becomes subdominant at small wavenumbers, where the behavior
is affected by gravitational collapse.
Therefore, the spectrum resembles that of a small-scale dynamo, where
the magnetic field is also in super-equipartition at large $k$;
see \cite{Hau+03,Hau+04}.
However, this similarity should not be regarded in any way as evidence
in favor of a dynamo.
It appears to be instead just a typical behavior of any type of
hydromagnetic turbulence.

Returning to the slight uprise of the magnetic field for $1\leq k/k_1<5$
in the case of an initially uniform magnetic field, we argue that this
is caused by the tangling of the magnetic field by the collapsing gas
motions.
It is therefore not due to an inverse cascade, which usually occurs
only in the absence of a mean magnetic flux through the domain.
In the case of a turbulent magnetic field, the build-up of vorticity
at small wavenumbers could be caused by the shear flows, which leads
to what is known as a vorticity dynamo \citep{EKR03}.
In the case of an initially uniform magnetic field,
this vorticity dynamo is suppressed \citep{KMB09}.
However, the uprise of the vortical part of the velocity field for
$1\leq k/k_1<5$ appears to be caused by the magnetic field and becomes
weaker for Runs~M2 and M3.

\subsection{Helicity spectra}

It is important to realize that, owing to the use of periodic boundary
conditions, an initially uniform magnetic field is equivalent to what
is sometimes described as an imposed magnetic field.
This is simply because the mean magnetic flux is preserved.
A well-known difference between cases with and without an imposed
magnetic field is the fact that $\bra{\AAA\cdot\BB}$ is no longer
conserved in the former case \citep{Berger97}.
This is because now the magnetic helicity in the domain interacts with
the magnetic helicity on scales larger than the size of the periodic
domain, but this part is no longer included in the simulation; see
the discussion in \cite{BM04}.
In simulations of decaying turbulence, it has been found that the magnetic
fluctuations decay more rapidly when there is an imposed magnetic field
\citep{Bran+20}.

\Fig{rspec_hel_comp} shows kinetic and magnetic helicity spectra for
the cases with a turbulent and an imposed field for runs with different
magnetic field strengths.
The kinetic helicity spectra are similar in the two cases, but the
magnetic helicity spectra are not.
In the case of an imposed magnetic field, there is magnetic helicity of
the same sign at all wavenumbers, although it is less strong at small
wavenumbers.
By contrast, in the case with a turbulent magnetic field with zero net
flux, the magnetic helicity is predominantly of opposite (negative) sign
and relatively strong also at small wavenumbers, except in the case with
the strongest magnetic field (Run~M1).
This is caused by magnetic helicity conservation, where a small-scale
driving of magnetic helicity of one sign causes automatically the
appearance of magnetic helicity of opposite sign at large scales; see
also \cite{Bran+19} for similar results.
The small-scale magnetic helicity does get slowly dissipated at late times
through finite microphysical magnetic diffusivity, leaving predominantly
the large-scale magnetic helicity of opposite sign behind.

\subsection{Contributions to the Lorentz work}

\begin{figure*}\begin{center}
\includegraphics[width=\textwidth]{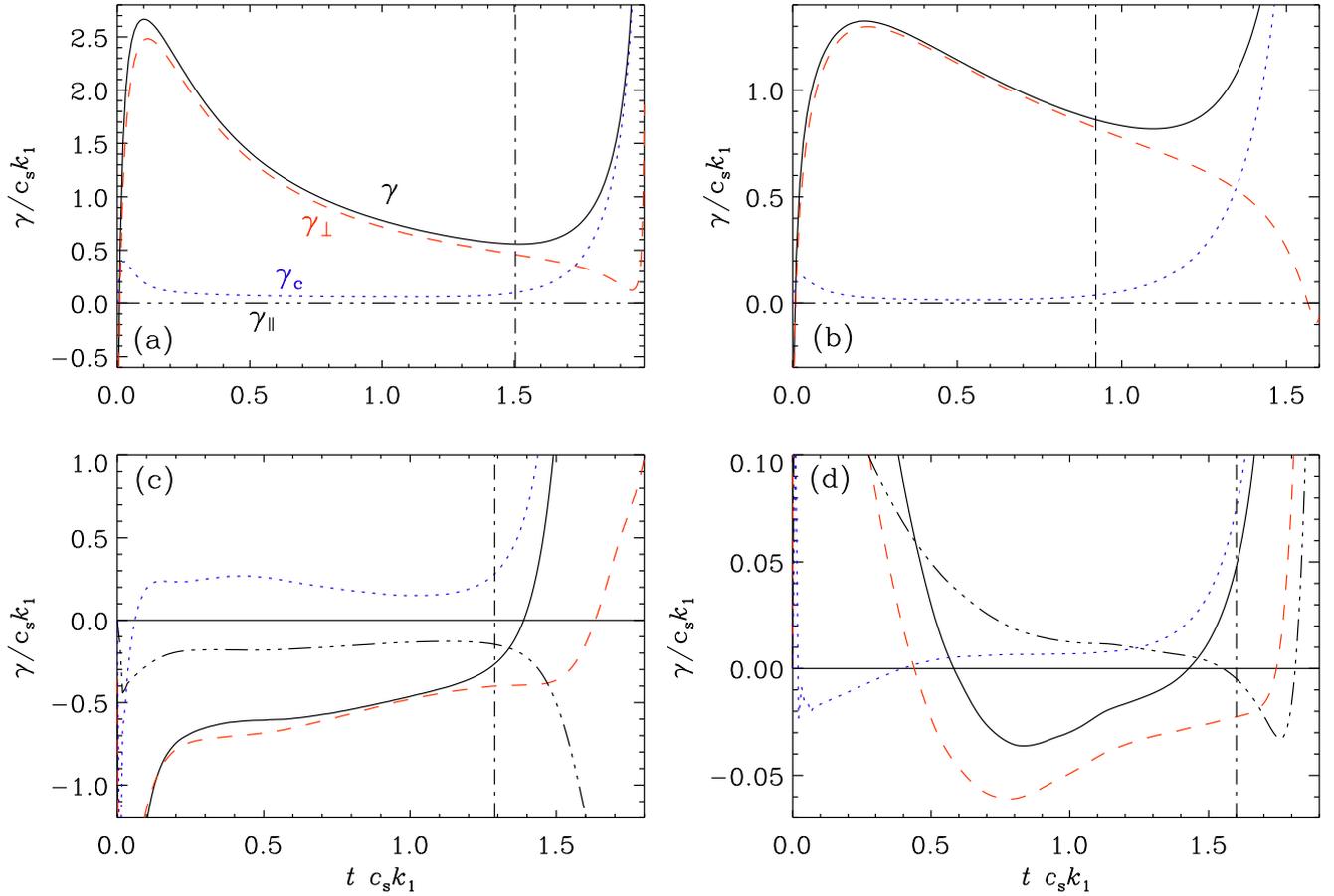}
\end{center}\caption[]{
Evolution of the pseudo growth rate
$\gamma=\gamma_{\rm c}+\gamma_\|+\gamma_\perp$ (solid lines),
with contributions from the work done against the curvature
force ($\gamma_\perp$, red dashed lines), the tension force
($\gamma_\|$, black dashed-triple-dotted lines), and the
magnetic pressure gradient ($\gamma_{\rm c}$, blue dotted lines)
for (a) Run~B, (b) Run~O2, (c) Run~M1, and (d) Run~I1.
In panels (a) and (b), $\gamma_\|=0$.
In panels (c) and (d), $\gamma_\|\neq0$,
and the zero line has been drawn as a straight black line.
The vertical dashed-dotted lines denote the critical time $t_*$ when
the Mach number has recovered to the original value of about 0.2.
}\label{pcomp_gammas}\end{figure*}

The work done against the Lorentz force serves as one of our main tools
to quantify dynamo action.
As we have mentioned at the end of \Sec{Energetics}, the work done against
the Lorentz force can be subdivided into contributions from the magnetic
pressure gradient, the tension force, and the curvature force.
In \Fig{pcomp_gammas}, we show that, for weak initial magnetic fields,
the most important contribution to the pseudo growth rate comes from the
work done against the curvature force, but later during the collapse,
a more important contribution comes from the compressional work done
against the magnetic pressure gradient.

\begin{figure*}\begin{center}
\includegraphics[width=\textwidth]{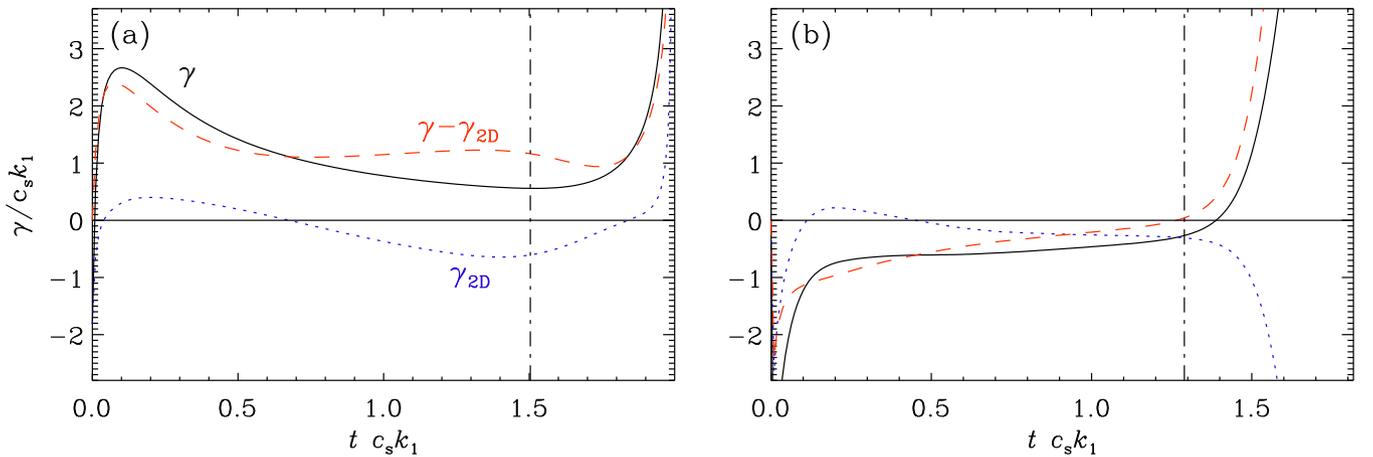}
\end{center}\caption[]{
Evolution of the pseudo growth rate $\gamma$ (black lines), with
contributions from $\gamma_{\rm 2D}$ (blue lines)
and the residual $\gamma-\gamma_{\rm 2D}$ (red lines),
for (a) Run~B and (b) Run~M1.
}\label{pWLcontri_comp}\end{figure*}

In the runs with a strong magnetic field (turbulent or imposed), the
value of $\gamma$ is still negative at the critical time $t_*$, but
the compressional work done against the magnetic pressure gradient is
positive, and it was positive also during the earlier phase, especially
in the case of a turbulent magnetic field; see \Fig{pcomp_gammas}(c).
The contribution of $\gamma_{\rm c}$ 
is a characteristic feature of amplification or at least sustenance
of the magnetic field in collapsing turbulence through compression.

In \Fig{pWLcontri_comp} we plot the time dependences of $\gamma$,
$\gamma_{\rm2D}$, and $\gamma_{\rm3D}=\gamma-\gamma_{\rm2D}$.
We see that $\gamma_{\rm2D}$ is always close to zero, except during an
early phase which can be associated with 2-D tangling of the initial
magnetic field.
When $\gamma_{\rm3D}$ is included, the resulting pseudo growth rate
is positive during much of the early part of the evolution.

Based on the positive values of $\gamma_\perp$ and $\gamma_{\rm3D}$ in the
cases of weak magnetic fields in \Figs{pcomp_gammas}{pWLcontri_comp}, we
are led to suggest that those runs do indeed host supercritical dynamos.
When the magnetic field is strong, however, $\gamma_\perp$ and $\gamma_{\rm3D}$
are now negative, suggesting that Run~M1 cannot be classified as a dynamo.

For strong magnetic fields, only near the end of the collapse does
$\gamma_\perp$ become positive.
On the other hand, $\gamma_{\rm3D}$ becomes then the dominant term
during the collapse and $\gamma_{\rm2D}$ becomes negative; see
\Figp{pWLcontri_comp}{b}.
This is probably caused by the strong alteration of the flow by the
magnetic field, making now $\gamma_{\rm2D}$ strongly negative.
This increases the compression and tangling terms associated with
$\gamma_{\rm2D}$, which then contribute to enhancing the kinetic
energy rather than the other way around (as in a dynamo).
Nevertheless, since the magnetic field is now increasing,
$\gamma_{\rm3D}$ becomes positive.
From \Figsp{pcomp_gammas}{c}{d} we know, however, that this
increase is manifestly caused by compression, which must therefore
be a 3-D compression, and not a dynamo effect.

In summary, we are led to conclude that there is dynamo action in
all cases with weak magnetic fields prior to collapse, but probably no
longer or not very much during the actual collapse.
When the magnetic field is already strong, there is no longer dynamo
action, but just 3-D compression.
For large Mach numbers, at late stages of the collapse, shocks form and
cross each other, which causes vorticity production \citep{Porter+15},
resulting then also in dynamo action \citep{Federrath+11}.
However, this happens at such small scales and such late times
that this effect cannot be captured in direct numerical simulations,
even at a resolution of $2048^3$ mesh points.
After $t\cs k_1=1.98$, the Jeans length is no longer resolved with 30 grid
cells -- the minimum proposed by \cite{Federrath+11b}; see \App{Density},
where we show the evolution of the maximum density during the collapse.

\begin{figure}\begin{center}
\includegraphics[width=\columnwidth]{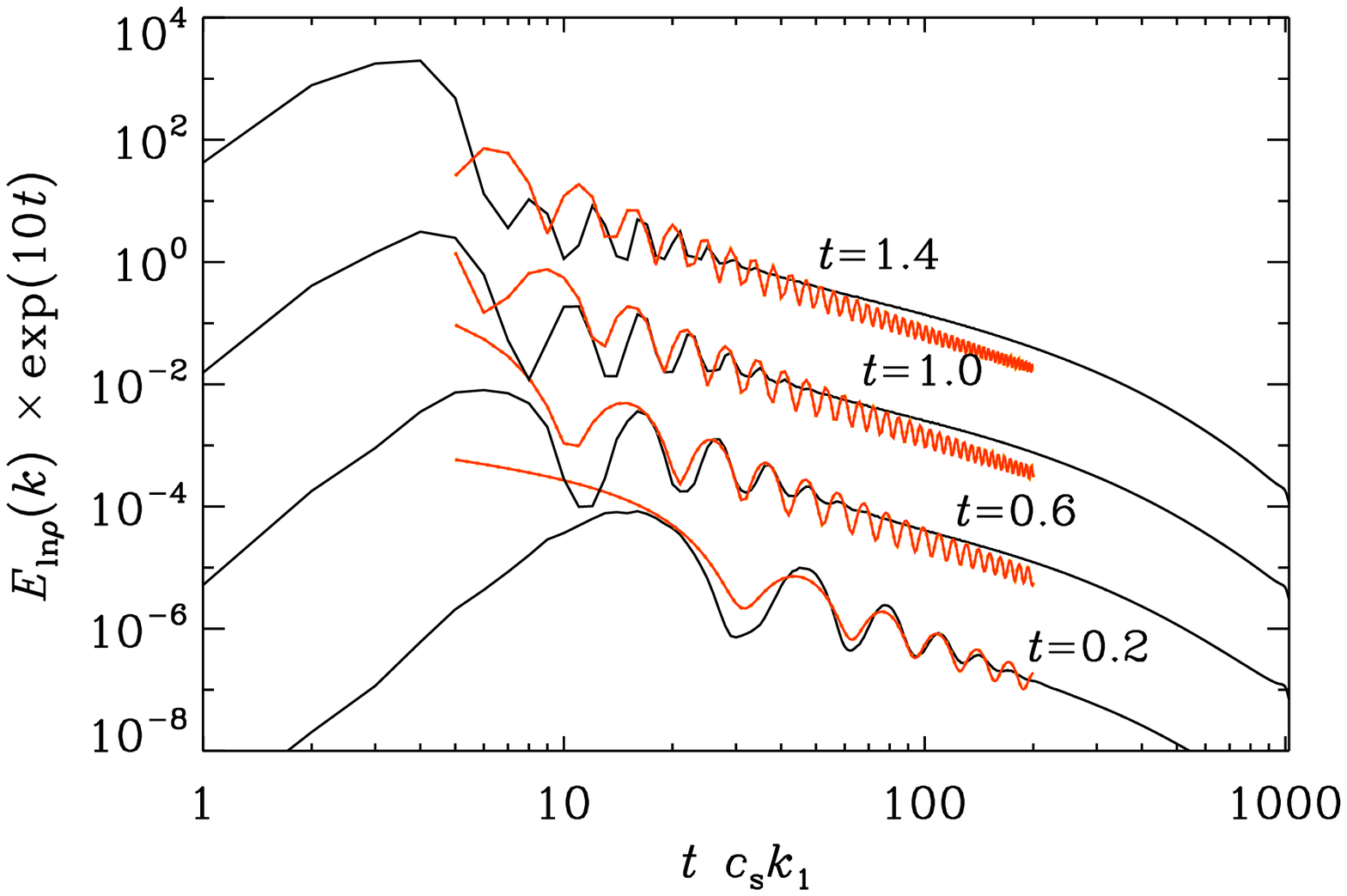}
\end{center}\caption[]{
Density spectra at $t=0.2$, 0.6, 1, and 1.4, shifted upward by an
$\exp(10t)$ factor to be able to see the shift of the oscillations
in the spectra in $k$.
The dotted orange lines denote approximate fits of the form given by
\Eq{ApproxFit}.
}\label{rspec_lr_S2048B_25_kf10f}\end{figure}

\subsection{Waves in the density spectra}

Before concluding, let us comment on an interesting feature that we
noticed in spectra of the logarithmic density.
At early and intermediate times, we see a wavy structure in
$E_{\ln\rho}(k)$; see \Fig{rspec_lr_S2048B_25_kf10f}.
In fact, this wavy modulation is of the form $\cos k\xi(t)$, where
$\xi(t)=\cs t$ is the distance a sound wave has propagated in the
time $t$ since the initial condition was applied.
As time goes on, and as $\xi(t)$ therefore increases, the waves appear
to propagate toward smaller values of $k$ and are of progressively shorter
length in $k$ space.
The changing phase of these waves can be described by the formula
\EQ
E_{\ln\rho}(k,t)=E_\rho^{(0)}(k)\,\left[1+g(k,t)\,(1-\cos k\cs t)\right],
\label{ApproxFit}
\EN
where $E_\rho^{(0)}(k)\propto k^{-5/3}$ denotes the unmodulated spectrum,
and $g(t)=(140/k)\exp(-0.8\,t)$ is an empirically defined function (in code units).
Only at later times, the fit is going somewhat out of phase.
Given the agreement of our hypothetical modulation of the form $\cos k\xi(t)$
with the actual spectrum, we can argue that the wavy structure is
indeed caused by the initial velocity perturbation launching sound waves
from multiple locations in the domain all at the same time, and that
their characteristic scale increases with time like $\xi(t)=\cs t$.
Similar waves have also been seen in simulations of gravitational
waves that are being initiated from an instantaneous perturbation;
see Figure~2 of \cite{RoperPol+20}.
No explanation for this phenomenon was offered there, but we have now
confirmed that it can be explained in a similar way, except that the
relevant speed in the expression for $\xi(t)$ is the speed of light.

\section{Conclusions}
\label{Conclusions}

In the present work, we have used the instantaneous excess of the work
done by the Lorentz force over the Joule dissipation as a quantity that
characterizes the dynamo.
Under stationary conditions, the dynamo can easily be characterized by
the growth rate.
However, a growth rate cannot be defined in situations when the velocity
itself decays or grows exponentially with time, like we observe in
our models.

The dynamo criterion based on the work terms results in a value of
the critical Reynolds number of about 25, which is smaller than the
critical value of about 35 for small-scale dynamo action \citep{Hau+04},
but larger than the value for large-scale dynamo action in the presence
of helicity of below six \citep{Bra09}.
Also, in the present case there is helicity, so we do expect a critical
value that is less than 35.
However, there is a strong contribution from irrotational motions that
makes the dynamo harder to excite and does itself not contribute to
dynamo action \citep{MB06}.
During the collapse, i.e., after $t=t_*$, and for weak magnetic fields
(Runs~B and O2), $\gamma_\perp$ and $\gamma_{\rm3D}$ show a slight
decrease when the magnetic field is weak, supporting the idea that this
magnetic field growth is not primarily caused by dynamo action, but just
by compression.

Our investigation shows that the most important contribution to the growth
of a magnetic field comes from the work done against the curvature force,
although later during the collapse, there is an even more important
contribution from the compressional work done against the magnetic
pressure gradient.
However, as we have argued above, this type of magnetic field
amplification happens also in one or two dimensions and should
therefore not be associated with dynamo action.
By considering the decomposition into $\gamma_{\rm2D}$ and $\gamma_{\rm3D}$
we have made an attempt of distinguishing dynamo action from
the type of non-dynamo amplification seen also in two dimensions.
Nevertheless, our dynamo criterion is not very precise, as the pseudo
growth rate changes behavior with different initial conditions and a
number of factors need to be considered in combination.

We stated in the introduction that the exact fraction of potential
energy that goes into turbulence is unknown.
Our results now show that one-third of the energy input from potential
energy goes into compressional heating, and two-third goes into the
kinetic and magnetic energies of the turbulence.
Thus, one would expect that, at the end of the collapse, the sum of
kinetic and magnetic energy densities is twice the thermal energy density
from compressive heating.
This is different from the virial theorem, which relates potential and
kinetic energies to each other.
As explained in \Sec{WorkTerms}, however, since the two-third contribution
to the kinetic energy comes from potential energy, which becomes more
negative with time, it follows that the ratio of kinetic to potential
energy is 2/3 and thus, the virial parameter is 4/3.
It would be unity, if the contribution to the kinetic energy was half
the Jeans work.
At later times, however, the fractional kinetic energy gain increases
toward 3/4 of the Jeans work, which implies a virial parameter of
about 3/2.

In all the simulations presented here, we have used an isothermal
equation of state. However, deviations from isothermality probably play
an important role during molecular cloud collapse.
For example, \citet{Lee+Hennebelle2018} caution that, in simulations
studying gravitational fragmentation of molecular clouds, an isothermal
equation of state cannot lead to converged results with increasing
numerical resolution.
They propose that an adiabatic equation of state at high densities,
essentially accounting for the formation of the Larson core is a more
physically meaningful approach.
In future work, it would be interesting to perform simulations using an
ideal gas equation of state instead of an isothermal one.
Such simulations could also allow for cooling, which would further
increase the density in regions of strong flow convergence and counteract
an otherwise singular collapse.

It would also be interesting to apply our analyses to earlier
work that uses Bonnor-Ebert spheres as initial conditions
\citep{Sur+10,Sur+12,Federrath+11b}.
Bonnor-Ebert spheres are nonuniform equilibria in a non-expanding finite space,
so there is no $\rho_0$ term in \Eq{del2Phi}.
Our simulations
never go through such a state or reach any near-equilibrium state.
Starting with a Bonnor-Ebert sphere could lead to the collapse
proceeding in a different way from ours.
If the collapse occurs sufficiently slowly,
it could be easier to achieve dynamo growth rates that
remain faster than the collapse rate for a longer time.

Another useful extension would be to compare with simulations that make
use of adaptive mesh refinement \citep[see, for example,][]{Federrath+10}.
Such simulations would have varying accuracy in space, and it is currently
unclear how this affects the kinetic and magnetic energy spectra and
other diagnostics.
Since the varying accuracy is not a concern in the present simulations,
they can be used as a benchmark.
Another advantage of the present simulations is the fact that the
viscosity and magnetic diffusivity are fixed and that therefore
numerically converged and accurate results are possible.

\section*{Acknowledgements}

We thank the referee for useful comments and references.
This work was supported in part through the Swedish Research Council,
grant 2019-04234.
EN acknowledges funding by the ERC Grant ``Interstellar''
(Grant agreement 740120) and the Hellenic Foundation for Research and
Innovation (Project number 224).
Nordita is supported in part by Nordforsk.
We acknowledge the allocation of computing resources provided by the
Swedish National Infrastructure for Computing (SNIC)
at the PDC Center for High Performance Computing Stockholm
and the National Supercomputer Centre (NSC) at Link\"oping.
\vspace{2mm}

\noindent
{\em Code and data availability.} The source code used for the
simulations of this study, the {\sc Pencil Code} \citep{JOSS},
is freely available on \url{https://github.com/pencil-code/}.
The DOI of the code is https://doi.org/10.5281/zenodo.2315093.
The simulation setup and the corresponding data are freely
available on \url{https://doi.org/10.5281/zenodo.5760126}; see also
\url{https://www.nordita.org/~brandenb/projects/SelfGrav} for easier
access to the same material as on the Zenodo site.

\appendix
\section{Virial parameter}
\label{Virial}

In \Sec{Energetics}, we noted that $\alpha_{\rm vir}=2\EEK/|{\cal E}_{\rm P}|$
is expected to be around unity, but that its value can be different at
large Mach numbers and for strong magnetic fields.
We have also stated that a value of 4/3 is expected if 2/3
of the Jeans work goes into building up kinetic energy.
The purpose of this appendix is now to compare the evolution of
$\alpha_{\rm vir}$ for runs with strong magnetic field (Run~M1), larger
Mach number (Run~S), with a subsonic run with weak magnetic field (Run~B).

\begin{figure}\begin{center}
\includegraphics[width=\columnwidth]{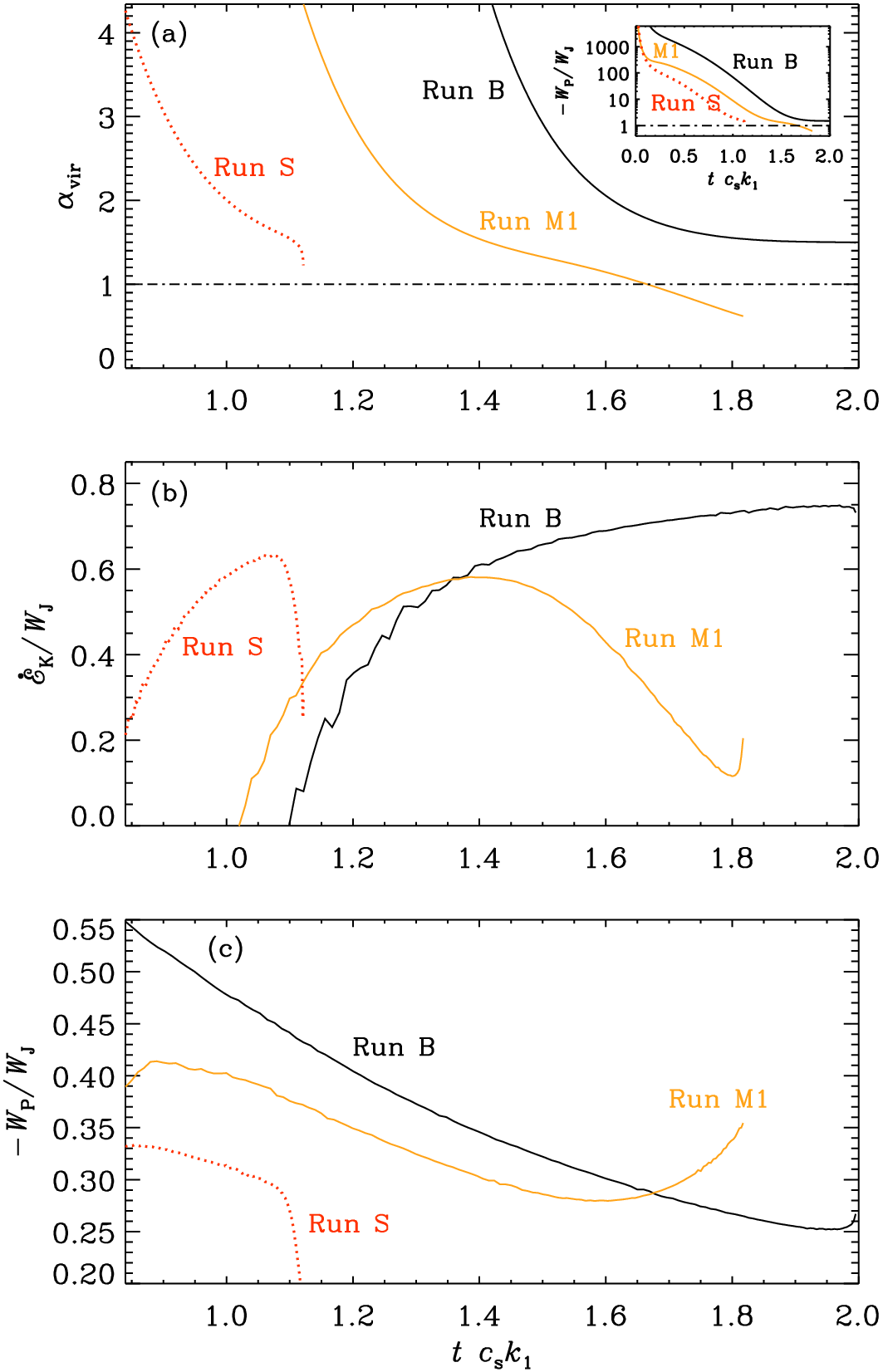}
\end{center}\caption[]{
Time evolution of (a) the virial parameter, (b) the fractional kinetic
energy gain, and (c) the fractional pressure work for Runs~B, M1, and S.
In (a), the inset is a logarithmic representation of $\alpha_{\rm vir}$
over a larger range.
At the end of Runs~S and M1, the results are affected by insufficient
resolution and cannot be trusted.
}\label{pepot}\end{figure}

Initially, when the density is uniform, the potential
energy density density $\EEP=-\bra{(\nab\Phi)^2}/8\pi G$
is small,\footnote{Using the identity
$\nab\cdot(\Phi\nab\Phi)=(\nab\Phi)^2+\Phi\nabla^2\Phi$, the potential
energy density can also be written as ${\cal E}_{\rm P}=-\bra{\rho\Phi}/2$,
where we have made use of periodicity and the fact that $\bra{\Phi}=0$.
This yields $\dot{\cal E}_{\rm P}=-\bra{\nab\Phi\cdot\nab\dot{\Phi}}/4\pi G
=-\bra{\dot{\rho}\Phi}$, which leads to \Eq{dEP} after using \Eq{Dlnrho}.}
so $\alpha_{\rm vir}$ is large.
However, when the collapse has started, deep potential wells develop
and $|{\cal E}_{\rm P}|$ increases with ${\cal E}_{\rm P}<0$, so
$\alpha_{\rm vir}$ drops and eventually settles at a value of around
$1.5$ for weak magnetic fields (Run~B), and perhaps also for
the supersonic run before the collapse occurred (Run~S); see \Figp{pepot}{a}.
This larger value $\alpha_{\rm vir}\approx1.5$ is caused primarily
by the fact that the fractional kinetic energy gain from the Jeans
work in \Eq{JeansRatios} increases with time from 2/3 to 3/4; see
\Figp{pepot}{b}.
At the same time, the fractional pressure work decreases correspondingly
from 1/3 to 1/4; see \Figp{pepot}{c}.
For strong magnetic fields (Run~M1), $\alpha_{\rm vir}$ 
continues to decrease below unity; see \Fig{pepot}.

\section{Growth of and relation between magnetic field and density}
\label{Growth}

\begin{figure}\begin{center}
\includegraphics[width=\columnwidth]{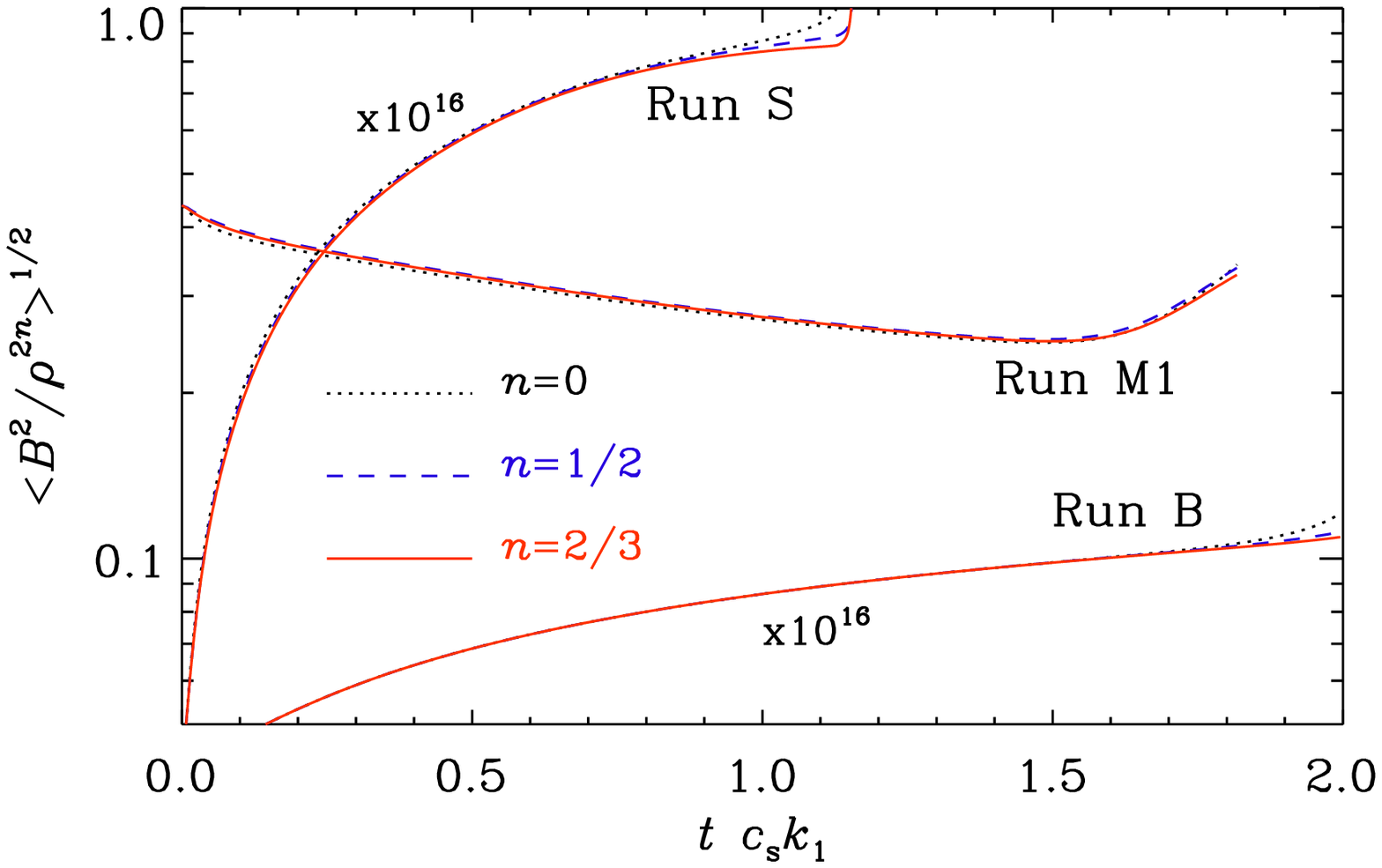}
\end{center}\caption[]{
Different scalings of the magnetic field strength
as a function of time (scaled up by $10^{16}$ for Runs~B and S).
Note that the units of $\bra{\BB^2/\rho^{2n}}^{1/2}$
are $\cs\mu_0^{1/2}\rho_0^{1/2-n}$.
}\label{pbdens}\end{figure}

\begin{figure}\begin{center}
\includegraphics[width=\columnwidth]{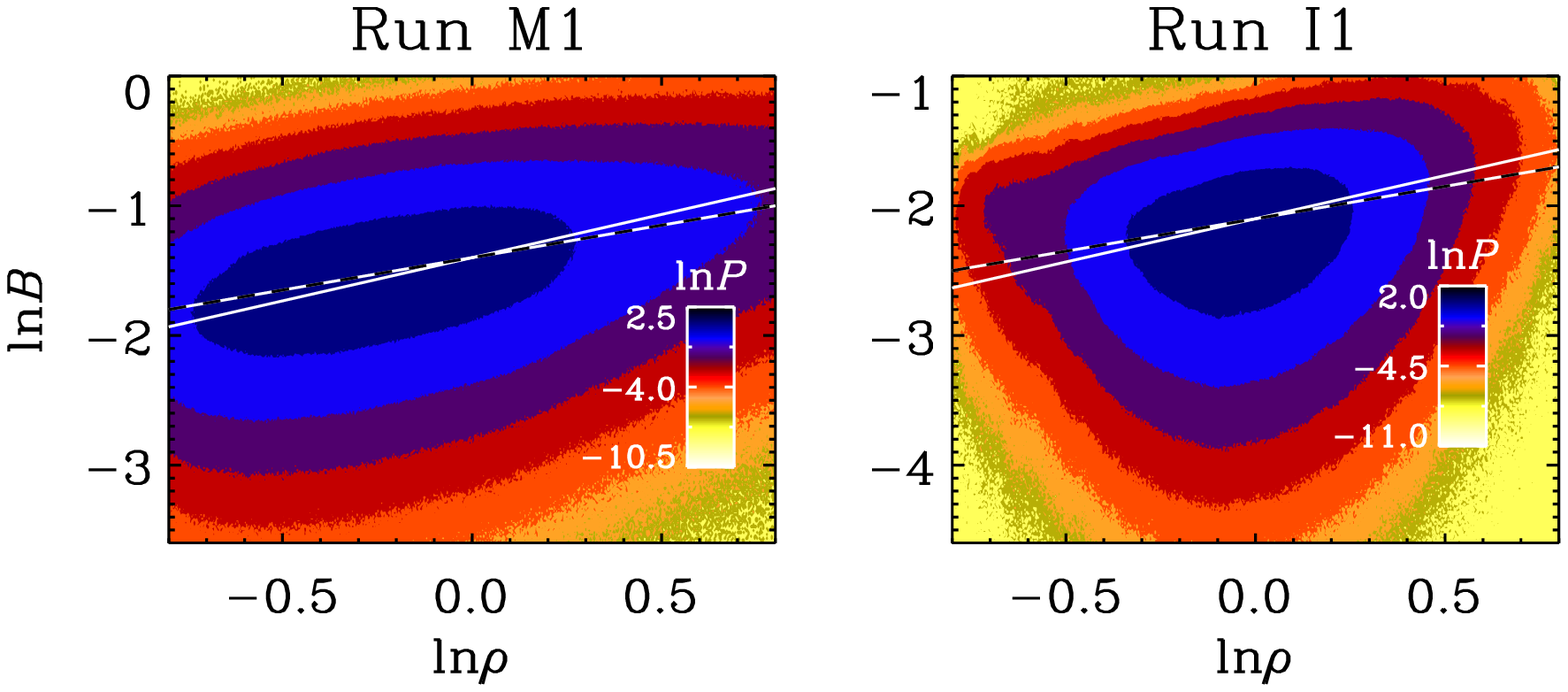}
\end{center}\caption[]{
Logarithm of 2-D histograms $P(\ln\rho,\ln|\BB|)$ for Runs~M1 and I1.
The solid and dashed lines correspond to $|\BB|\sim\rho^{2/3}$
and $|\BB|\sim\rho^{1/2}$, respectively.
}\label{ppdf2d_comp_last}\end{figure}

\begin{figure}\begin{center}
\includegraphics[width=\columnwidth]{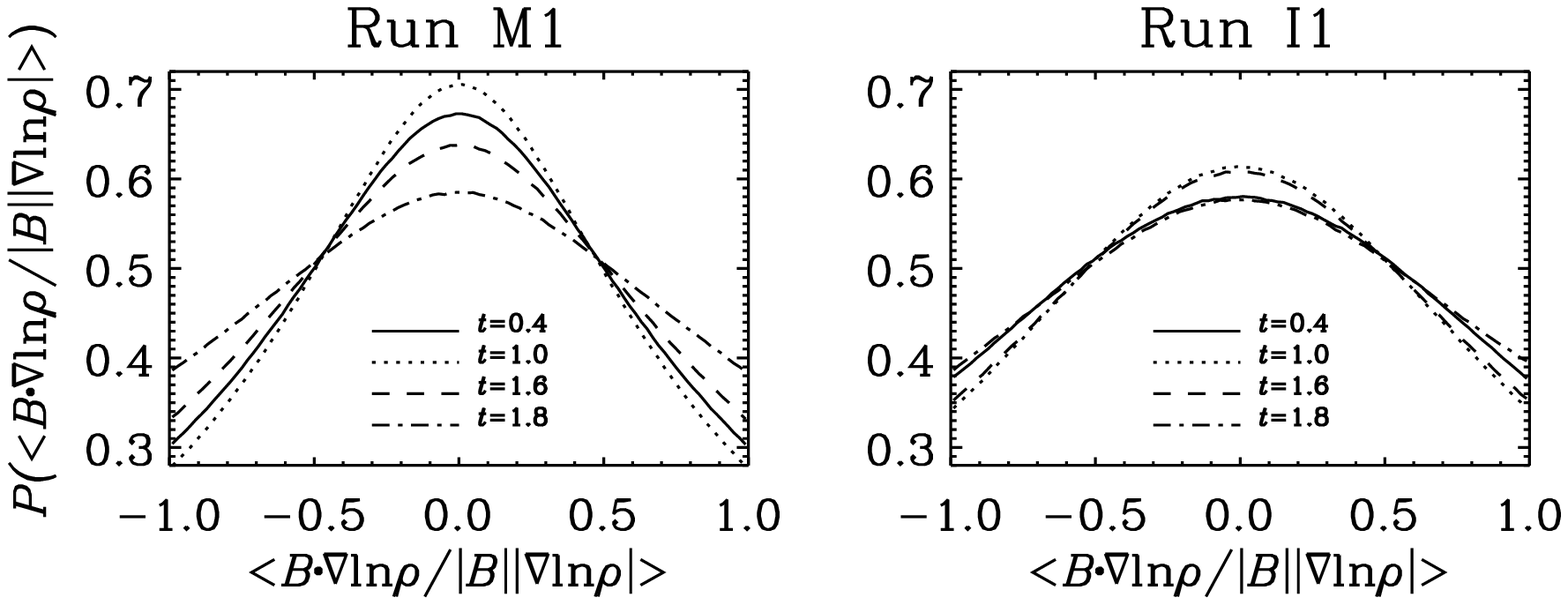}
\end{center}\caption[]{
Histograms $P(\BB\cdot\nab\ln\rho/|\BB||\nab\ln\rho|)$ for Runs~M1 and I1,
for different times.
}\label{ppdf_comp}\end{figure}

The purpose of this appendix is to assess the role of the density in
determining a relevant measure of the magnetic field in collapsing
turbulence.
During radial collapse, a uniform magnetic field is amplified such
that the ratio $|\BB|/\rho^{2/3}$ is constant \citep{Pudritz+06}.
Therefore, it is customary to monitor the evolution of this
quantity \citep{Sur+10,Sur+12,Federrath+11b,Sharda+21}.
As a suitable volume average, one can consider
$|\BB|/\rho^{2/3}\sim \bra{\BB^2/\rho^{4/3}}^{1/2}$.
In \Fig{pbdens}, we show that for Runs~B and M1, the quantities
$\bra{\BB^2/\rho^{2n}}^{1/2}/\rho_0^{n}$ are nearly the same for
different exponents $n$.
For Run~S, the differences for different $n$ are somewhat larger, but the
differences between the cases $n=1/2$ and $n=2/3$ are still negligible.
This justifies the use of $\vA^{\rm rms}$ in \Fig{pcomp_urms2} of the
main text, which corresponds to $n=1/2$ in \Fig{pbdens}.
In fact, also the case $n=1/2$ has been discussed previously in the
context of gravitational collapse \citep{Crutcher1999}.
In \Fig{ppdf2d_comp_last}, we present the logarithm of 2-D histograms
$P(\ln\rho,\ln|\BB|)$ to show that most of the points in the volume lie
within elliptical islands stretched along the line $|\BB|\sim\rho^{2/3}$.
They are normalized such that
$\int P(\ln\rho,\ln|\BB|)\,\dd\ln\rho\,\dd\ln|\BB|=1$.
This behavior has been observed in numerous simulations of
self-gravitating turbulence \citep[e.g.][]{soler_2013,chen2016,Barreto-Mota2021},
as a result of the converging motions driven by gravity
\citep{soler_hennebelle2017}.

In \Sec{Visualizations}, we mentioned that $\ln\rho$ lacks small-scale
structure, and that only $\nab\ln\rho$ displays noticeable small-scale
variations.
In \Fig{ppdf_comp} we present histograms of the cosine of
the angle between $\BB$ and the logarithmic density gradients,
$P(\BB\cdot\nab\ln\rho/|\BB||\nab\ln\rho|)$, for Runs~M1 and I1 at
different times.
They show that $\BB$ is mostly perpendicular to $\nab\ln\rho$, i.e., the
magnetic field lines tend to be aligned with the contours of $\ln\rho$.

\section{Comparison with the Taylor microscale Reynolds number}
\label{Relam}

\begin{figure}\begin{center}
\includegraphics[width=\columnwidth]{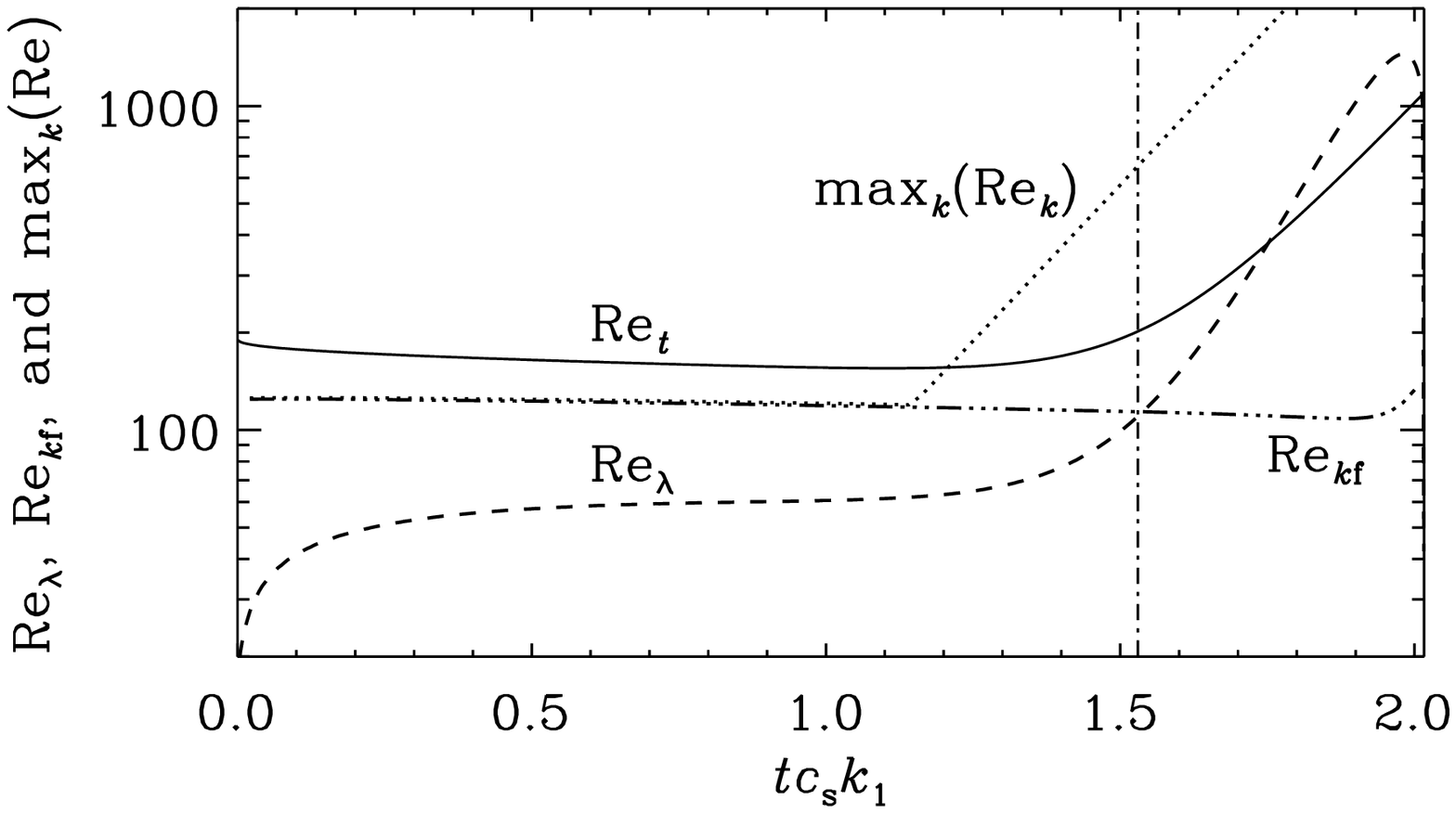}
\end{center}\caption[]{
Comparison of the evolution of the Taylor microscale Reynolds number
with other Reynolds numbers for Run~B.
}\label{rlam}\end{figure}

In \Fig{rlam}, we compare the evolution of the Taylor microscale Reynolds
number, $\Rey_\lambda$, with other Reynolds numbers: $\Rey_{\kf}$
and $\Rey_t$, whose values at $t=t_*$ are given in \Tab{Tsummary} for
$\kf/k_1=10$, and the maximum of $\Rey_{k}$ over $k$, $\max_k(\Rey_k)$,
which is at later times dominated by the peak at small $k$; see
\Fig{rspec_last}, where the peak is at $k/k_1\approx2$.
Note that $\max_k(\Rey_k)$ begins to grow exponentially
shortly after $t\cs k_1\approx1.1$.

In \Sec{Conclusions}, we compared our critical magnetic Reynolds number
of 25 with the earlier value of 35 by \cite{Hau+04b}, and we argued
that the new value is smaller because of helicity in the flow.
We also stated that the results depend on the Mach number \citep{Hau+04b}.
We can now compare with the highly compressible, supersonic hydromagnetic
turbulence simulations of \cite{Federrath+14}, who gave a critical
magnetic Reynolds number of $\approx130$.
This value is based on half the size of the computational domain.
To convert it to the normalization of \cite{Hau+04,Hau+04b}, it
should be divided by $2\pi$.
This results in a value of $\approx20$, which is smaller than the values
of 35 to 70 found by \cite{Hau+04b} for Mach numbers below and above
unity, respectively.
However, when they increased their magnetic Prandtl number from unity
to five, they found $\Rmc=25$ and $50$ for Mach numbers below and above
unity, respectively; compare Figs.~7 and 8 of \cite{Hau+04b}.
\cite{Federrath+14} used $\Pm=10$, so their value of $\Rmc=20$ cannot
directly be compared with those of \cite{Hau+04b} for $\Pm=5$, but it
seems a bit low.
This question cannot be fully clarified here and might depend on subtle
numerical aspects.
The {\tt FLASH} code \citep{Fryxell+00} used by \cite{Federrath+14}
is based on a Riemann solver, which may affect the effective viscosity
and magnetic diffusivity.

\section{Maximum density and other density moments}
\label{Density}

\begin{figure}\begin{center}
\includegraphics[width=\columnwidth]{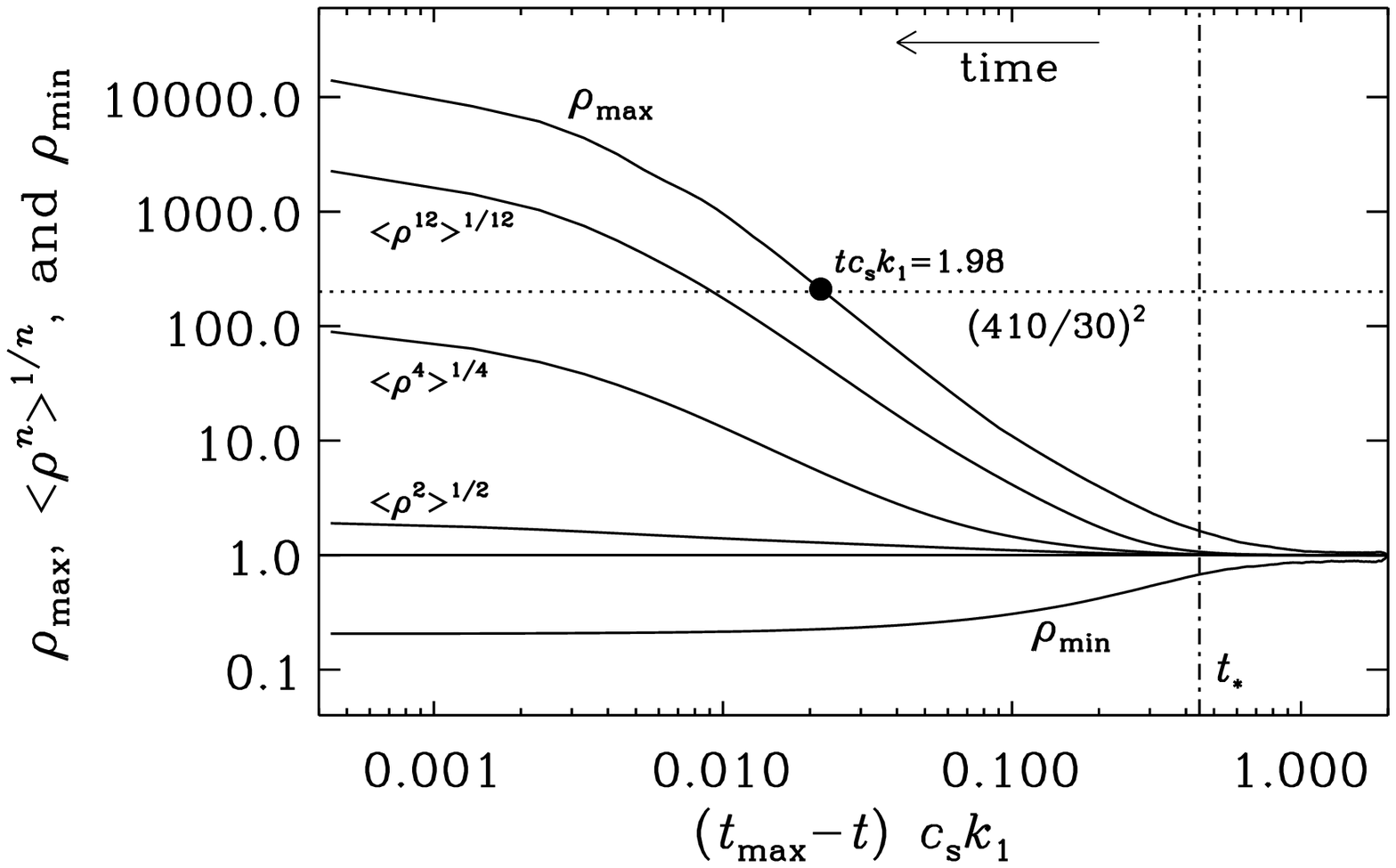}
\end{center}\caption[]{
Evolution of the maximum density along with the minimum density and
intermediate moments of the density for $n=2$, $4$, and $12$.
Note that time increases toward the left.
The time $t\cs k_1=1.98$ for which the Jeans length becomes resolved by
less than 30 mesh points is marked by the black symbol.
This is when $(\rho_{\max}/\rho_0)^{1/2}=410/30$.
}\label{pdensity}\end{figure}

As time goes on and the collapse proceeds, the density contrast increases.
This has implications for the nominal Jeans length, which decreases with
increasing maximum density.
To get an idea of this, we plot in \Fig{pdensity} the evolution of the
maximum density.
For completeness, we also plot the minimum density and intermediate
moments of the density, $\bra{\rho^n}^{1/n}$ for $n=2$, $4$, and $12$.
Note that
\EQ
\rho_{\max}=\lim_{n\to\infty}\bra{\rho^n}^{1/n},
\quad\mbox{and}\quad
\rho_{\min}=\lim_{n\to-\infty}\bra{\rho^n}^{1/n}.
\EN
To show the values close to the collapse time $t_{\max}$ more clearly, we
plot the densities versus $(t_{\max}-t)\,\cs k_1$ in a doubly-logarithmic
representation.

The nominal Jeans length is proportional to $\rho^{-1/2}$, although
the standard stability analysis breaks down if the density is no longer
constant.
We see that the resolution criterion of \cite{Federrath+11b}
is reached when the square root of the density exceeds 1/30
of the initial Jeans length of 410 grid cells for
$\sigma_{\rm J}/\cs k_1=5$; see \Sec{Units}.

\bibliographystyle{mnras}
\bibliography{ref}{}
\end{document}